\documentclass[usenatbib,useAMS]{mn2e}
\usepackage{times}
\usepackage{amssymb}
\usepackage{epsfig}
\usepackage{color}

\renewcommand{\vec}[1]{\bmath{#1}}
\newcommand*\myhrulefill{%
\leavevmode\leaders\hrule depth-1.9pt height 2.0pt\hfill\kern0pt
}

\begin{document}
\title[OGLE-2008-BLG-510 --  weak microlensing anomaly]{OGLE-2008-BLG-510: first automated real-time detection of a weak microlensing anomaly
-- brown dwarf or stellar binary?\thanks{based in part on data collected by MiNDSTEp with
the Danish 1.54m telescope at the ESO La Silla Observatory}}
\author[V. Bozza et al.]{V. Bozza$^{1}$, M. Dominik$^{2}\thanks{E-mail: {\tt md35@st-andrews.ac.uk}}$\thanks{Royal Society University Research Fellow}, N. J. Rattenbury$^{3}$,
U. G. J{\o}rgensen$^{4,5}$, Y. Tsapras$^{6,7}$, \newauthor
D. M. Bramich$^{8}$, A. Udalski$^{9}$, I. A. Bond$^{10}$, C. Liebig$^{2,11}$, A. Cassan$^{11,12}$
% ...
\\
$^{1}$Universit\`{a} degli Studi di Salerno, Dipartimento di Fisica "E.R.~ Caianiello", Via S.~Allende, 84081 Baronissi (SA), Italy\\
$^{2}$SUPA, University of St Andrews, School of Physics \&
Astronomy, North Haugh, St Andrews, KY16 9SS, United Kingdom\\
$^{3}$Department of Physics, University of Auckland, Private Bag 92-019, Auckland 1001, New Zealand\\
$^{4}$Niels Bohr Institute, University of Copenhagen, Juliane Maries Vej 30, 2100 Copenhagen, Denmark\\
$^{5}$Centre for Star and Planet Formation, Geological Museum,
   {\O}ster Voldgade 5-7, 1350 Copenhagen, Denmark\\
$^{6}$Las Cumbres Observatory Global Telescope Network,
6740B Cortona Dr, Goleta, CA 93117, United States of America \\
$^{7}$Astronomy Unit, School of Mathematical Sciences, Queen Mary, University of London, London E1 4NS, United Kingdom\\
$^{8}$ESO Headquarters, Karl-Schwarzschild-Str. 2, 85748 Garching bei M\"{u}nchen, Germany \\
$^{9}$Warsaw University Observatory, Al. Ujazdowskie 4, 00-478 Warszawa, Poland\\
$^{10}$Institute for Information and Mathematical Sciences, Massey University, Private Bag 102-904, Auckland 1330,
New Zealand\\
$^{11}$Astronomisches Rechen-Institut, Zentrum f\"{u}r Astronomie der Universit\"{a}t Heidelberg (ZAH),  M\"{o}nchhofstr.\ 12-14,
69120 Heidelberg, Germany\\
$^{12}$Institut d'Astrophysique de Paris, 75014, Paris, France\\
% ...
{\rm \normalsize (continued on last page)}}

\maketitle

\clearpage



\begin{abstract}
The microlensing event OGLE-2008-BLG-510 is characterised by an evident asymmetric shape of the peak, promptly detected by the
ARTEMiS system in real time. The skewness of the light curve appears to be compatible both with binary-lens and binary-source models, including the possibility that the lens system consists of an M dwarf orbited by a brown dwarf. The detection of this microlensing anomaly and our analysis demonstrates that: 1) automated real-time detection of weak microlensing anomalies with immediate feedback is feasible, efficient, and sensitive, 2) rather common weak features intrinsically come with ambiguities that are not easily resolved from photometric light curves, 3) a modelling approach that finds all features of parameter space rather than just the `favourite model' is required, and 4) the data quality is most crucial, where systematics can be confused with real features, in particular small higher-order effects such as orbital motion signatures. It moreover becomes apparent that events with weak signatures are a silver mine for statistical studies, although not easy to exploit. Clues about the apparent paucity of both brown-dwarf companions and binary-source microlensing events might hide here.
\end{abstract}


\section{Introduction}


The 'most curious' effect of gravitational microlensing \citep{Ein36,Pac86} lets us extend our knowledge of planetary systems \citep{MP91} to a region of parameter space unreachable by other methods and thus populated with intriguing surprises. 
Microlensing has already impressively demonstrated its sensitivity to Super-Earths with the detection of a 5 Earth-mass (uncertain to a factor two) planet \citep{PLANET:planet}, and it reaches down even to about the mass of the Moon \citep{PacRev}.

The transient nature of microlensing events means that rather than the characterisation of individual systems, it is the population statistics that will provide the major scientific return of observational campaigns. Meaningful statistics will however only arise with a controlled experiment, following well-defined fully-deterministic and reproducible procedures. In fact, the observed sample is a statistical representation of the true population under the respective detection efficiency of the experiment. 
An analysis of 13 events with peak magnifications $A_0 \geq 200$ observed between 2005 and 2008 provided the first well-defined sample \citep{Gould10}. In contrast, the various claims of planetary signatures and further potential signatures come with vastly different degrees of evidence and arise from different data treatments and applied criteria \citep{Do:Gerg} as well as observing campaigns following different strategies. 

While the selection of highly-magnified peaks is relatively easily controllable, and these come with a particularly large sensitivity to planetary companions to the lens star \citep{GriSaf,Horne2025}, their rarity poses a fundamental limit to planet abundance measurements. Moreover, the finite size of the source stars strongly disfavours the immediate peak region for planet masses $\la 10~M_{\oplus}$, where a large magnification results from source and lens star being very closely aligned. In contrast, during the wing phases of a microlensing event, planets are more easily recognised with larger sources because of an increased signal duration, as long as the amplitude exceeds the threshold given by the photometric accuracy \citep[c.f.\ ][]{Han:priorities}. It is therefore not a surprise at all that the two least massive planets found so far with unambiguous evidence from a well-covered anomaly, namely OGLE-2005-BLG-390Lb \citep{PLANET:planet} and MOA-2009-BLG-266Lb \citep{Muraki:planet}, come with an off-peak signature at moderate magnification with a larger source star.

An event duration of about a month and a probability of $\sim\,10^{-6}$ for an observed Galactic bulge star to be substantially brightened at any given time \citep{Pac91,KP94} called for a 2-step strategy of survey and follow-up observations \citep{GL92}. 
In such a scheme, surveys monitor $\ga 10^{8}$ stars
on a daily basis for ongoing microlensing events \citep{OGLE,MACHO,MOA,EROS:bulge}, whereas roughly hourly sampling of the most promising ongoing events with a network of telescopes supporting round-the-clock coverage and photometric accuracy of $\la 2\,\%$ allows not only the detection of planetary signatures, but also their characterisation \citep{PLANET1,PLANET:EGS,RoboNet-1.0}. While real-time alert systems on ongoing anomalies \citep{OGLE:alert,MACHO:alert}, combined with the real-time provision of photometric data, paved the way for efficient target selection by follow-up campaigns, the real-time identification of planetary signatures and other deviations from ordinary light curves ('anomalies')
\citep{OGLE:EEWS,SIGNALMEN} allowed a transition to a 3-step-approach, where the regular follow-up cadence can be relaxed in favour of monitoring more events, and a further step of anomaly monitoring at $\sim\,$5--10~min cadence (including target-of-opportunity observations) is added, suitable to extend the exploration to planets of Earth mass and below.

The recent and upcoming increase of the field-of-view of microlensing surveys (MOA: $2.2~\rmn{deg}^2$, OGLE-IV: $1.4~\rmn{deg}^2$, Wise Observatory: $1~\rmn{deg}^2$, KMTNet: $4~\rmn{deg}^2$) allows for sampling intervals as small as 10--15~min. 
This almost merges the different stages with regard to cadence, but the surveys are to choose the exposure time (determining the photometric accuracy) per field rather than per target star. Moreover, they cannot compete with the angular resolution possible with lucky-imaging cameras \citep{Lucky1,Lucky2,SONG:Uffe,SONG}, given that this technique is incompatible with a wide field-of-view. This leaves a most relevant role for ground-based follow-up networks in breaking into the regime below Earth mass, in particular with space-based surveys \citep{Bennett:space} at least about a decade away. 

It is rather straightforward to run microlensing surveys in a fully-deterministic way, but it is very challenging to achieve the same for both the target selection process of follow-up campaigns and the real-time anomaly identification. 
The pioneering use of robotic telescopes in this field with the
RoboNet campaigns \citep{RoboNet-1.0,RoboNet-II} led \citet{Horne2025} to devise the first workable target prioritisation algorithm. While the OGLE EEWS \citep{OGLE:EEWS} was the first automated system to detect potential deviations from ordinary microlensing light curves, it flagged such suspicions to humans who would then take a decision. In contrast, the {\sc signalmen} anomaly detector \citep{SIGNALMEN} was designed just to rely on statistics and request further data from telescopes 
until a decision for or against an ongoing anomaly can be taken with sufficient confidence. {\sc signalmen} has already demonstrated its power by detecting the first sign of finite-source effects in MOA-2007-BLG-233/OGLE-2007-BLG-302 \citep{Choi:forthcoming} -- ahead of any humans -- and leading to the first alerts sent to observing teams that resulted in crucial data being taken on OGLE-2007-BLG-355/MOA-2007-BLG-278 \citep{Han:several} and OGLE-2007-BLG-368/MOA-2007-BLG-308 \citep{Sumi:abundance}, the latter involving a planet of $\sim\,20~M_{\oplus}$.
{\sc signalmen} is now fully embedded into the ARTEMiS (Automated Robotic Telescope Exoplanet Microlensing Search) software system \citep{ARTEMiS1,ARTEMiS2} for data modelling and visualisation, anomaly detection, and target selection. 
The 2008 MiNDSTEp (Microlensing Network for the Detection of Small Terrestrial Exoplanets) campaign directed by ARTEMiS provided a proof of concept for fully-deterministic follow-up observations  \citep{MiNDSTEp0}.

In event OGLE-2008-BLG-510, discussed in detail here, evidence for an ongoing microlensing anomaly was for the first time obtained by an automated feedback loop realised with the ARTEMiS system, without any human intervention.
This demonstrates that robotic or quasi-robotic follow-up campaigns can operate efficiently. 

A fundamental difficulty in obtaining planet population statistics arises from the fact  that many microlensing events show weak or ambiguous anomalies, sometimes with poor quality data, which are left aside because the time investment in their modelling would be too high compared to the dubious perspective to draw any definite conclusions. Indeed such
events represent a silver mine for statistical studies yet to be designed. Event OGLE-2002-BLG-55 has already been very rightfully classified as ''a possible planetary event'' \citep{OGLE55,Gaudi55}, where ambiguities are mainly the result of sparse data over the suspected anomaly. Here, we show that OGLE-2008-BLG-510 makes another example for ambiguities, which in this case arise due to the lack of prominent features of the weak perturbation near the peak of the microlensing event. Given that $\chi^2$ is not a powerful discriminator, in particular in the absence of proper noise models \citep[e.g.\ ][]{Ansari}, it needs a careful analysis of the constraints on parameter space posed by the data rather than just a claim of a 'most favourable' model. In fact, the latter might point to excitingly exotic configurations, but it must not be forgotten that maximum-likelihood estimates (equalling to $\chi^2$ minimisation under the assumption of measurement uncertainties being normally distributed) are not guaranteed to be anywhere near the true value. 
We therefore apply a modelling approach that is based on a full classification of the finite number of morphologies of microlensing light curves in order to make sure that no feature of the intricate parameter space is missed (Bozza et al., in preparation).

\citet{Do:binary} argued that the apparent paucity of microlensing events reported that involve source rather than lens binaries \citep[e.g.\ ][]{GriHu} could be the result of an intrinsic lack of characteristic features, but despite a further analysis by \citet{Han:binary}, the puzzle is not solved yet. Moreover, while all estimates of the planet abundance from microlensing observations indicate a quite moderate number of massive gas giants \citep{Sumi:abundance, Gould10, Do:abundance,PLANET:abundance}, the small number of reported brown dwarfs \citep[c.f.\ ][]{Do:Gerg}, much easier to detect, seems even more striking. As we will see, the case of OGLE-2008-BLG-510 appears to be linked to both. Maybe the full exploration of parameter space for events with weak or without any obvious anomaly features will get us closer to understanding this issue which is of primary relevance for deriving abundance statistics.


In Sect.~2, we report the observations of OGLE-2008-BLG-510 along with the record of the anomaly detection process and the strategic choices made. 
Sect.~3 details the modelling process, whereas the competing physical scenarios are discussed in Sect.~4, before we present final conclusions in Sect.~5.



\section{Observations and anomaly detection}

\begin{table*}
\begin{center}
\caption{Overview of campaigns that monitored OGLE-2008-BLG-510 and the telescopes used}
\label{tab:campaigns}
\begin{tabular}{clll}
\hline
Campaign & Telescope & Site & Country\\
\hline
OGLE & Warsaw 1.3 m  & Las Campanas Observatory (LCO) & Chile \\
MOA & MOA 1.8m & Mt John University Observatory (MJUO) & New Zealand \\
MiNDSTEp & Danish 1.54m & ESO La Silla & Chile \\
MicroFUN & SMARTS 1.3m & Cerro Tololo Inter-American Observatory (CTIO) & Chile \\
RoboNet-II & Faulkes North 2.0m (FTN)$^{\star}$ & Haleakala Observatory & United States (HI)\\
\ldots & Faulkes North 2.0m (FTS)$^{\star}$ & Siding Spring Observatory (SSO) & Australia (NSW)\\
\ldots & Liverpool 2.0m (LT) & Observatorio del Roque de Los Muchachos & Spain (Canary Islands) \\
PLANET-III & Elizabeth 1.0m & South African Astronomical Observatory (SAAO) & South Africa \\
 \ldots   & Canopus 1.0m & Canopus Observatory, University of Tasmania & Australia (TAS) \\
 \ldots  & Lowell 0.6m & Perth Observatory & Australia (WA) \\
 (ToO) & IRSF 1.4m & South African Astronomical Observatory (SAAO) & South Africa \\
  (ToO) & SALT 11m  & South African Astronomical Observatory (SAAO) & South Africa \\
\hline
\end{tabular}
\end{center}
\begin{flushleft}
OGLE: Optical Gravitational Lensing Experiment, MOA: Microlensing Observations in Astrophysics,
MiNDSTEp: Microlensing Network for the Detection of Small Terrestrial Exoplanets, PLANET: Probing Lensing Anomalies NETwork,
(ToO): target-of-opportunity observations at further sites not participating in regular microlensing observations.\\
$^\star$ The FTN and FTS are part of the Las Cumbres Observatory Global Telescope (LCOGT) Network
\end{flushleft}
\end{table*}

\begin{table*}
\begin{center}
\caption{Timeline of OGLE-2008-BLG-510 observations and anomaly detection}
\label{tab:timeline}
\begin{tabular}{ccp{12cm}}
\hline
Date & Time & \\
\hline
28 July 2008 & 15:12 UT & Event OGLE-2008-BLG-510 announced by OGLE \\
3 August 2008 & & Event selected by the ARTEMiS system for MiNDSTEp follow-up observations \\
4 August 2008 & 0:50 UT & First MiNDSTEp data from the Danish 1.54m at ESO La Silla (Chile) \\
                        & 1:52 UT & First MicroFUN data from the CTIO 1.3m (Chile) \\
7 August 2008 & & First RoboNet-II data from the Faulkes North 2.0 (FTN, Hawaii), Faulkes South 2.0m (FTS, Australia), and Liverpool 2.0m (LT, Canary Islands)\\
8 August 2008 & 0:39 UT & First PLANET-III data from the SAAO 1.0m (South Africa)\\
9 August 2008 & 1:30 UT & Event announced as MOA-2008-BLG-369 by MOA, following independent detection\\
{\bf \ldots} & {\bf 4:58 UT} & {\sc signalmen}  suspects anomaly based on Danish 1.54m data, acquired at 4:41 UT; 7 further data points were taken with this telescope lead to revision of model parameters (peak 8 hours later)\\
\ldots & 11:43 UT & PLANET-III starts observing the event with the Canopus 1.0m of the University of Tasmania\\
{\bf \ldots} & {\bf 21:01 UT} & {\sc signalmen} suspects anomaly based on SAAO 1.0m data acquired at 20:28 UT, and with 3 further SAAO data points available by 22:00 UT, the model parameters were revised again (peak expected another 8 hours later) \\
{\bf 10 August 2008} & {\bf 5:39 UT} &  {\sc signalmen} triggers check status based on a data point acquired with the Danish 1.54m at 5:36 UT (just before dawn) \\
{\bf \ldots} & {\bf 6:01 UT} & Deviation confirmed by further data promptly taken data at 5:47, 5:50, and 5:55 UT, prompting to a stronger rise than expected, but {\sc signalmen} 1 data point short of calling an anomaly \\
\ldots & 9:09 UT & Reassessment triggered by OGLE data point obtained at 3:35 UT, found to be deviating, but final assessment unchanged\\
{\bf \ldots} & {\bf 9:12 UT} & The MiNDSTEp and ARTEMiS teams circulate an e-mail to all other microlensing observing teams pointing to an ongoing anomaly\\
{\bf \ldots} & {\bf 13:00 UT} & {\sc signalmen} evaluation of 4 data points taken as part of the regular MOA observations between 8:08 UT and 10:50 UT led to the automated activation of `anomaly' status. As a result, further data were taken with the telescopes already observing the event, and moreover the IRSF 1.4m and the SALT, both at SAAO (South Africa). \\

\hline
\end{tabular}
\end{center}
\begin{flushleft}
(boldface highlights epochs marked in Fig.~\protect\ref{Fig LCpart})
\end{flushleft}
\end{table*}

The microlensing event OGLE-2008-BLG-510 at RA~18$^\rmn{h}$09$^\rmn{m}$37$\fs$65 and Dec~$-26^\circ$02$\arcmin$26$\farcs$70  (J2000), first discovered by the Optical Gravitational Lensing Experiment (OGLE), was subsequently monitored by several campaigns with telescopes at various longitudes (see Table~\ref{tab:campaigns}), and independently detected as  MOA-2008-BLG-369 by the Microlensing Observations in Astrophysics (MOA) team. Table~\ref{tab:timeline} presents a timeline of observations and anomaly detection 


When the follow-up observations started, the event magnification was estimated by {\sc signalmen} \citep{SIGNALMEN} to be $A \sim 3.9$, which for a baseline magnitude $I\sim 19.23$ and absence of blending means an observed target magnitude $I \sim 17.75$. The event magnification implied an initial sampling interval for the MiNDSTEp campaign of $\tau = 60~\rmn{min}$ according to the MiNDSTEp strategy \citep{MiNDSTEp0}.

The OGLE, MOA, MiNDSTEp, RoboNet-II, and PLANET-III groups all had real-time data reduction pipelines running, and with efficient data transfer, photometric measurements were available for assessment of anomalous behaviour by {\sc signalmen} shortly after the observations had taken place. While the MicroFUN team also took care of timely provision of their data, we did not manage at that time to get a data link with {\sc signalmen} installed, but since 2009 we enjoy an efficient {\sc rsync} connection.\footnote{{\sc rsync} is a software application and network protocol for synchronising data stored in different locations that keeps data transfer to a minimum by efficiently working out differences. --- {\tt http://rsync.samba.org}} RoboNet data processing unfortunately had to cease due to fire in Santa Barbara in early July. After resuming of operations and working through the backlog, RoboNet data on OGLE-2008-BLG-510 were not available before 23 August 2008.

The {\sc signalmen} anomaly detector \citep{SIGNALMEN} is based on the principle that real-time photometry and flexible scheduling allow requesting further data for assessment until a decision about an ongoing anomaly can be taken with sufficient confidence. The specific choice of the adopted algorithm comes with substantial arbitrariness, where the power for detecting anomalies needs to be balanced carefully against the false alert rate. {\sc signalmen} assigns a "status" to each of the events, which is either 'ordinary' (= there is no ongoing anomaly), 'anomaly' (= there is an ongoing anomaly), or 'check' (= there may or may not be an ongoing anomaly). This "status" triggers a respective response: 'ordinary' events are scheduled according to the standard priority algorithm, 'anomaly' events are alerted upon, initially monitored at high cadence, and given manual control for potentially lowering the cadence, while for 'check' events further data at high cadence are requested urgently until the event either moves to 'anomaly' or back to 'ordinary' status.

{\sc signalmen} also adopts strategies to achieve robustness against problems with the data reduction and to increase sensitivity to small deviations, namely the use of a robust fitting algorithm that automatically downweights outliers and its own assessment of the scatter of reported data rather than reliance on the reported estimated uncertainties. As a result, {\sc signalmen} errs on the cautious side by avoiding to trigger anomalous behaviour on data with large scatter at the cost of missing potential deviations. The {\sc signalmen} algorithm has been described by \citet{SIGNALMEN} in every detail. We just note here that suspicion for a deviant data point that gives rise to a suspected anomaly is based on fulfilling two criteria, which asymptotically coincide for normally distributed uncertainties: (1) the residual is larger than 95\,\% of all residuals, (2) the residual is larger than twice the median scatter. This implies that {\sc signalmen} is expected to elevate events to 'check' status for about 5\,\% of the incoming data, but the power of detecting anomalies outweighs the effort spent on false alerts. In order to allow a proper evaluation of the scatter, for each data set, at least 6 data points {\em and} observations from at least 2 previous nights are required. {\sc signalmen} moves from 'check' to 'anomaly' mode with a sequence of at least 5 deviant points found.

On event OGLE-2008-BLG-510, there were two early suspicions of an anomaly, both on 9 Aug 2008 (see Table~\ref{tab:timeline}. 
Each of these lead to a revision of the model parameters with the event expected about 8 hours later than estimate earlier, rather than finding conclusive evidence of an ongoing anomaly. In fact, this behaviour is indicative of the event flattening out its rise earlier than expected. Moreover, weak anomalies over longer time-scales look marginally compatible with ordinary events at early stages, and failure to match expectations can result in a series of deviations that let {\sc signalmen} trigger 'check' mode, which then leads to a revision of the model parameters rather than to a firm detection of an anomaly, only to happen once stronger effects become evident. 

On 10 August 2008, {\sc signalmen} predicted a magnification of $A \sim 15.6$, which meant a sampling interval of $30~\mathrm{min}$ for the MiNDSTEp observations with the Danish 1.54m. Since the event was first alerted by OGLE, with a peak magnification 
$A_0 = 4.6 \pm 13.3$,\footnote{The quoted uncertainty refers to a locally linearised model.}
the {\sc signalmen} estimate had changed from initially $A_0 \sim 2.3$ to $A_0 \sim 5.3$ when the first data with the Danish 1.54m were obtained to now
 $A_0 \sim 17$ (which bears some similarity with `model 1c', discussed in the next section). In contrast to a maximum-likelihood estimate which tends to overestimate the peak magnification, the maximum a posteriori estimate used by {\sc signalmen} tends to underestimate it \citep{Albrow:MAP,SIGNALMEN}. Just before dawn in Chile, an ongoing anomaly was suspected again, with further data subsequently leading to firm evidence. Despite increased airmass likely affecting the measurements towards the end of the night, the earlier 'check' triggers were indicative of a real anomaly being in progress.



MOA, Canopus 1.0m, and FTN were able to cover the peak region, which looked evidently
asymmetric. The descent was covered by SAAO and then by the Danish 1.54m. Unfortunately, the Moon heavily disturbed the
observations in the descent, inducing large systematic errors that
were difficult to correct in the offline reduction. Furthermore,
we missed three nights of MOA data (12-14 August), five nights of
OGLE (11-15 August) and we had a two-day hole between 12 and 14
August not covered by any telescopes. Follow-up observations were
resumed after 14 August, including PLANET-III observations with the Perth 0.6m (Western Australia)
from 15 to 20 August, and continued until 30 August. After then,
only the OGLE and MOA survey telescopes continued to collect data.

We found all data being affected by a scatter larger than what
might be reasonably expected by the error bars assigned by the
reduction software. In fact, following the anomaly around
the peak of the light curve, all telescopes suffered from inferior data
quality resulting from increased sky flux contribution due to the proximity
of the full Moon to the target field as well as contamination by a nearby
saturated star. The relative faintness of the microlensed source ($I=19.2$ according to OGLE database)
added to the problems. A re-analysis of the images obtained from the FTN, LT, FTS, Danish 1.54m, SAAO 1.0m, and Canopus 1.0m
telescope using {\sc DanDIA}\footnote{{\sc DanDIA} is built from the {\sc DanIDL} library of {\sc IDL}
routines available at {\tt http://www.danidl.co.uk}} \citep{bramich08} made a crucial difference by removing previously present systematics which could have easily been mistaken as indications of higher-order effects, and posed a puzzle in the interpretation of this event. {\sc DanDIA} is an 
implementation of the difference imaging technique \citep{alard98} whose specific power arises from modelling the kernel as a discrete pixel array, allowing us to properly deal with distorted star profiles because it makes no underlying assumptions
about the shape of the point-spread function.

\begin{figure*}
\centering{ \resizebox{16cm}{!}{\includegraphics{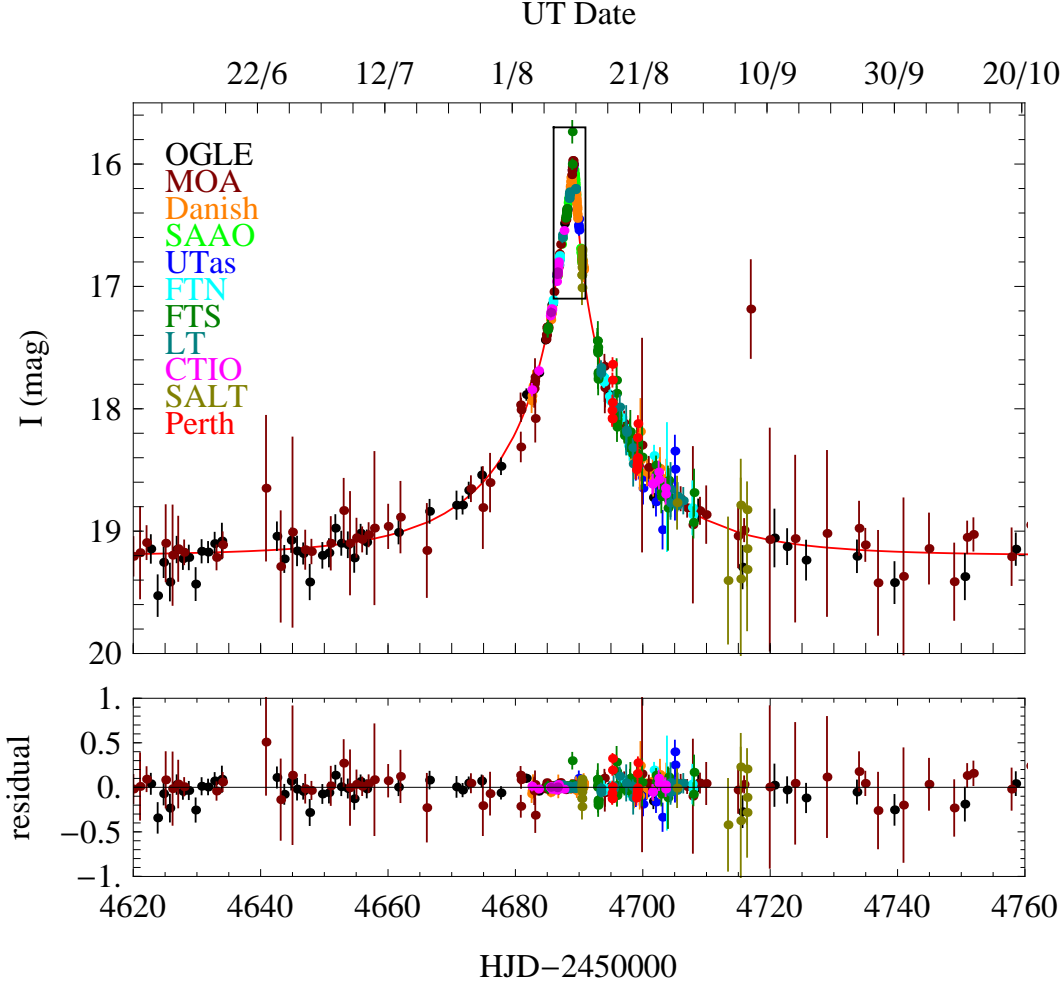}}}
 \caption{Data acquired with several telescopes (colour-coded) on gravitational microlensing event OGLE-2008-BLG-510 (MOA-2008-BLG-369) along with the best-fitting model light curve and the respective residuals. The region marked by the box is expanded in Fig. \ref{Fig LCpart}.}
 \label{Fig LC}
\end{figure*}

\begin{figure*}
\centering{ \resizebox{16cm}{!}{\includegraphics{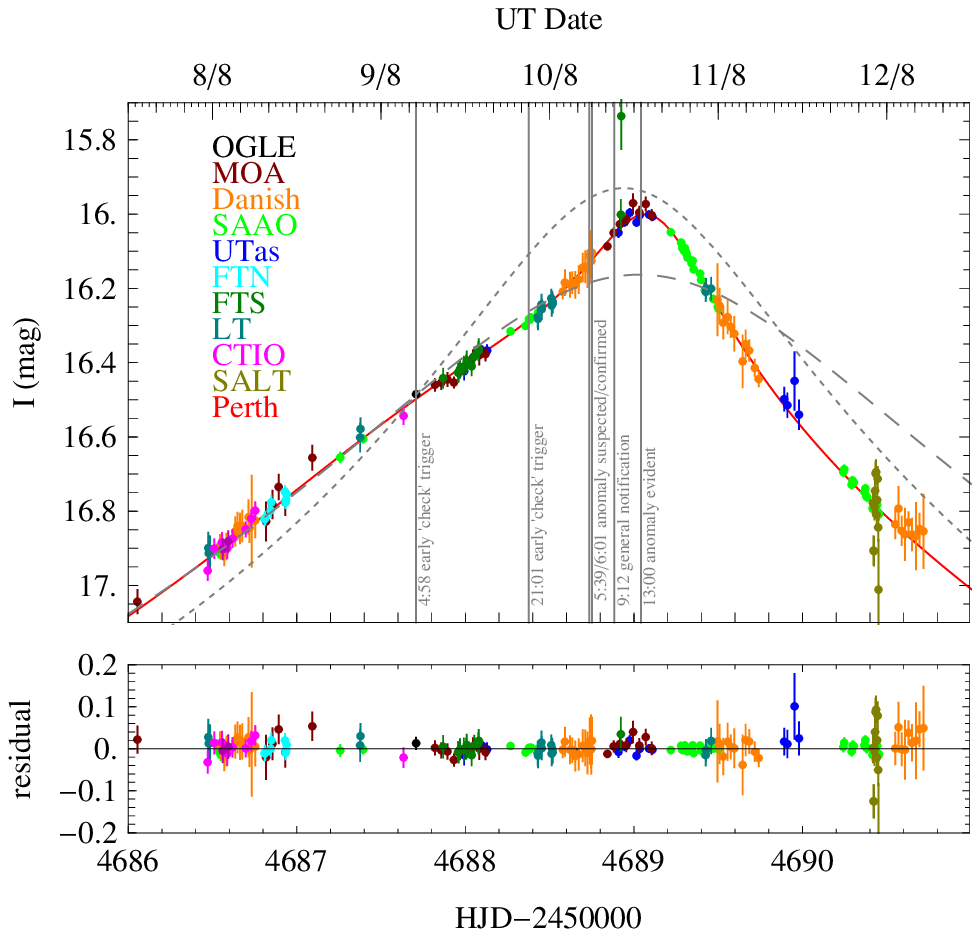}}}
 \caption{Data and model light curve for OGLE-2008-BLG-510 around the asymmetric peak. For comparison, we also show best-fitting model light curves for a single lens star using all data (short dashes) or data before triggering on the anomaly on 10 August 2008 (long dashes). Moreover, we have indicated the most relevant stages in the real-time identification of the anomaly.}
 \label{Fig LCpart}
\end{figure*}

In our analysis we have used the data taken at all mentioned
telescopes except for CTIO, IRSF, SALT, and Perth 0.6m, since they are too sparse or too scattered to significantly
constrain the fit. Moreover, we have neglected all data prior to $\mathrm{HJD} = 2454500$, where the light curve is flat and
therefore insensitive to the model parameters. Furthermore, we have re-binned most of the data taken
in the nights between 12 and 16 August disturbed by the Moon since they were very scattered and redundant. Finally, we have
re-normalised all the error bars so as to have $\chi^2$ equalling the number of degrees of freedom 
for the model with lowest $\chi^2$, which is related to the {\em assumption} that it provides a reasonable explanation of the observed data. These prepared data sets are shown in Fig. \ref{Fig LC} along with the best-fitting model that
will be presented and discussed in detail in Section 4. The peak anomaly is illustrated in more detail in Fig.
\ref{Fig LCpart}.

Looking at the data and the model light curve, it seems in fact that the early triggers on 9 August 2008, spanning the region
$4687.69 \leq \mathrm{HJD} - 2450000.0 \leq 4688.39$ were due to a real anomaly, but its weakness together with the limited photometric precision and accuracy did not allow us to obtain sufficient evidence. 

\section{Modelling}

\subsection{Microlensing light curves}

Gravitational microlensing events show a transient brightening of an observed source star that results from the gravitational bending of its light by an intervening object, the `gravitational lens'. The gravitational microlensing effect of a body with mass $M$ is characterised by the angular Einstein radius 
\begin{equation}
\theta_\rmn{E}=\sqrt{\frac{4GM}{c^2}\,\frac{\pi_\rmn{LS}}{1~\mbox{AU}}}\,,
\label{eq:thetaE}
\end{equation}
where $G$ is the universal gravitational constant, $c$ is the vacuum speed of light, and
\begin{equation}
\pi_\mathrm{LS} = 1~\mbox{AU}\,\left(D_\rmn{L}^{-1}-D_\rmn{S}^{-1}\right)
\end{equation}
is the relative parallax of the lens and source stars at distances $D_\rmn{L}$ and $D_\rmn{S}$ from the observer, respectively.

With source and lens star separated by an angle $u\,\theta_\rmn{E}$ on the sky, the magnification becomes \citep{Ein36}
\begin{equation}
A(u) = \frac{u^2+2}{u\,\sqrt{u^2+4}}\,.
\end{equation}
If we assume a uniform relative proper motion $\mu$ between lens and source star, the separation parameter $u$ becomes
\begin{equation}
u(t;t_0,u_0,t_\rmn{E}) = \sqrt{u_0^2 + \left(\frac{t-t_0}{t_\rmn{E}}\right)^2}\,,
\end{equation}
where $t_\rmn{E} = \theta_\rmn{E}/\mu$ is the event time-scale, and the closest angular approach $u_0$ is realised at time $t_0$.

With $F_\rmn{S}$ being the flux of the observed target star, and $F_\rmn{B}$ the background flux, the total observed flux becomes
\begin{equation}
F(t) = F_\rmn{S} A(t) + F_\rmn{B} = F_\rmn{base} \frac{A(t)+g}{1+g} =  F_\rmn{base} A_\rmn{obs}(t)
\end{equation}
with the baseline flux $F_\rmn{base} = F_\rmn{S} + F_\rmn{B}$ and the blend ratio $g = F_\rmn{B}/F_\rmn{S}$,
where
\begin{equation}
A_\rmn{obs}(t) =  \frac{A(t)+g}{1+g}
\end{equation}
is the observed magnification.

Because of $A(u)$ monotonically increasing as $u \to 0$, ordinary microlensing light curves (due to single point-like source and lens stars) are symmetric with respect to the peak at $t_0$, where the closest angular approach between lens and source $u(t_0) = u_0$ is realised, and fully characterised by $(t_0,u_0,t_\rmn{E})$ and the set of $(F_\rmn{S},F_\rmn{B})$ or $(F_\rmn{base},g)$ for each observing site and photometric passband. Best-fitting $(F_\rmn{S},F_\rmn{B})$ follow analytically from linear regression, whereas the observed flux is non-linear in all other parameters.

\subsection{Anomaly feature assessment and parameter search}
Apparently, the only evident feature pointing to an anomaly in OGLE-2008-BLG-510 is the asymmetric shape of the peak (see Fig~\ref{Fig LCpart}). Such weak effects can be explained by a finite extension of the central point caustic of a single isolated lens star due to binarity (which includes the presence of an orbiting planet). Moreover, the absence of strong effects excludes the source star from hitting or passing over the caustic. 
This straightforwardly restricts parameter space, and one could readily identify a limited number of viable configurations with regard to the topologies of the caustic and the source trajectory and exclude all others. The use of an 'event library' where the most important features are stored had already been suggested by \citet{MaoDiStef}, while generic features were the starting points for the exploration of parameter space in early efforts of modelling microlensing events \citep{Do:MLMC1,Do:Fits}. Moreover, the identification of features for caustic-passage events has been proven powerful in efficient searches for all viable configurations \citep{CFalgo,Cassanpar,Kains}.
More recent work built upon the universal topologies of binary-lens systems \citep{Erdl} in order to classify light curves \citep{Bozza:classification,Night08}. Based on the earlier work by \citet{Bozza:classification}, we adopted an automated approach (Bozza et al., in preparation) that starts from 76 different initial conditions covering all possible caustic crossings and cusp approaches in all caustic topologies occurring in binary lensing (close, intermediate and wide) \citep{Erdl,Do:ambiguity}. From these initial conditions, we have run a Levenberg-Marquardt algorithm for downhill fitting, and then we have refined the $\chi^2$ minima by Markov chains.


The roundish shape of the peak however does not allow us to immediately dismiss the alternative hypothesis that the source rather than the lens is a binary system. Binary-source light curves are simply the superposition of ordinary light curves \citep{GriHu}, leading to significant zoo of morphologies, which is however less diverse than that of binary lenses. Binary-lens systems can be uniquely identified from slope discontinuities and the sharp features that are associated with caustics, while smooth, weakly perturbed light curves may be ambiguous, and even potential planetary signatures might be mimicked by binary-source systems  \citep{Gaudi:source}.

\subsection{Binary-lens models}

\begin{figure}
\centering{\resizebox{\hsize}{!}{\includegraphics{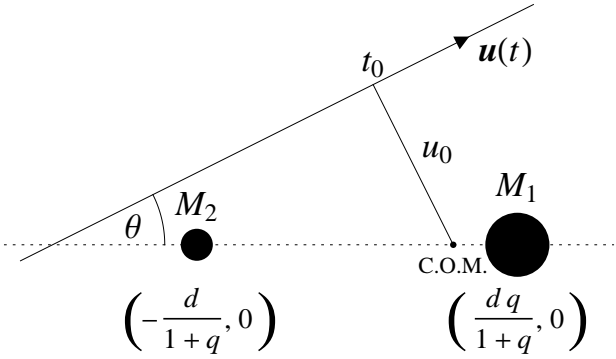}}}
 \caption{The geometry of a binary gravitational lens being approached by a single source star and the related parameters. The separation between the primary and the secondary is given by $d$, while the mass ratio is $q = M_2/M_1$. Positions on the sky arise from multiplying the dimensionless coordinates with the angular Einstein radius $\theta_\rmn{E}$, so that $t_\rmn{E} = \theta_\rmn{E}/\mu$ given an event time-scale, where $\mu$ denotes the absolute value of the relative proper motion between source and lens star.}
 \label{fig:BLparameters}
\end{figure}

In addition to its total mass $M = M_1+M_2$, a binary lens is fully characterised by the mass ratio $q = M_2/M_1$ of its constituents, and, if one neglects the orbital motion, by their separation. The latter can be described by the dimensionless parameter $d$, where $d\,\theta_\rmn{E}$ is the angle on the sky between the primary and the secondary as seen from the observer. In contrast to a single lens, the microlensing light curve depends on the orientation of the source trajectory, where we measure the angle $\theta$ from the axis pointing from $M_2$ to $M_1$. As the reference point for the closest angular approach between lens and source, characterised by $t_0$ and $u_0$, we choose the centre of mass. This means that the source trajectory relative to the lens is described by
\begin{equation}
\vec{u}(t) = u_0 \left( \begin{array}{c} -\sin \theta \\ \cos \theta \end{array} \right) + \frac{t-t_0}{t_\rmn{E}}\,\left( \begin{array}{c} \cos \theta \\ \sin \theta \end{array}\right)\,,
\end{equation}
while the primary of mass $M_1$ is at the angular coordinate $[d\,q/(1+q),0]\,\theta_\rmn{E}$ and the secondary of mass $M_2$ at  the angular coordinate $[-d/(1+q),0]\,\theta_\rmn{E}$ (see Fig.~\ref{fig:BLparameters}).

Strong differential magnifications also result in effects of the finite size of the source star on the light curve, quantified by the 
dimensionless parameter $\rho_\star$, where $\rho_\star\,\theta_\rmn{E}$ is the angular source radius. We initially approximate the star as uniformly bright. 



For the evaluation of the magnification for given model parameters, we have adopted a
contour integration algorithm \citep{Do:contour1,GG:contour,Do:contour2} improved with parabolic correction, optimal sampling and accurate error estimates, as described in detail by \citet{Bozza:contour}.

\begin{figure}
\centering{\resizebox{\hsize}{!}{\includegraphics{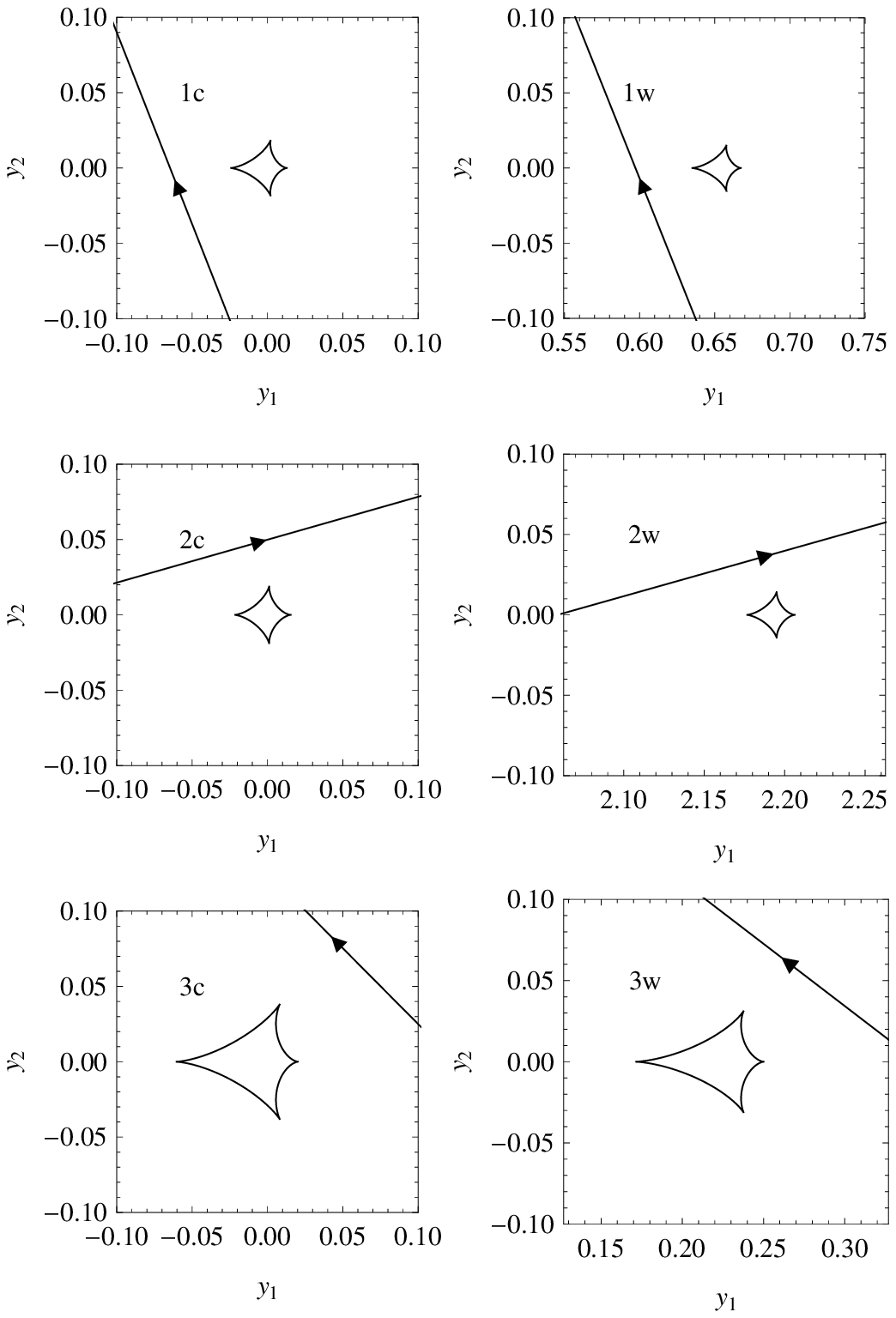}}}
 \caption{Illustration of the viable caustic and source trajectory configurations, showing how the trajectory approaches a cusp and passes with respect to the caustic. Models 1 and 2 differ in the exchanged roles of the on-axis and off-axis cusp, whereas models 2 and 3 exchange the on-axis cusp. Geometry and symmetry would allow for a 4th class of models, where the closest cusp approach is to the on-axis cusp on the side opposite the companion, but such were found not to be viable. }
 \label{Fig static}
\end{figure}

Our morphology classification approach leads to three viable configurations where the source trajectory approaches the caustic near the primary with different orientation angles, grazing one of the four cusps before having passed a neighbouring cusp at larger distance. We recover the well-known ambiguity between close and wide binaries \citep{GriSaf,Do:ambiguity}: all configurations come in two flavours. Fig.~\ref{Fig static} illustrates these configurations, labelled `1', `2', and `3', for the close-binary topology, whereas the wide-binary case is analogous. Given the symmetry of the binary-lens system with respect to the binary axis, there are 3 different cusps, one off-axis and two on-axis. For the off-axis cusp, there are two different neighbouring cusps, distinguishing models 2 and 3. In contrast, the neighbouring cusps to an on-axis cusp is the identical off-axis cusp, so that model 1 is not doubled up. In principle, there could be a solution near the approach of the other on-axis cusp, but this turned out not to be viable.


We end up with six candidate models that we label by 1c, 2c, 3c, 1w, 2w, 3w. The "c" corresponds to the close-binary topology and the "w" corresponds to the dual wide-binary topology. The values of the parameters of these models, with their uncertainties are shown in Table~\ref{Tab:static}. Model 1c comes with the smallest $\chi^2$ (set to the number of degrees of freedom by rescaling the photometric uncertainties), with model 1w closely following with just $\Delta \chi^2 = 1$. Models 2 and 3 come with $\Delta \chi^2 \sim 20$, with model 2w being singled out by its wide parameter ranges.
Models 1 prefer an OGLE blend ratio close to zero, whereas models 2 prefer a larger background, but are still compatible with zero blending. Models 3 come with a negative blend ratio. If one imposes a non-negative blend ratio, models 1 and 2 change rather little (see Table~\ref{Tab:staticplus}), but we did not find corresponding minima for the configurations of models 3, which rather tend towards models 2. Mass ratios $q$ roughly span the range from 0.1 to 1. Given that the source passes too far from the caustic to provide significant finite-source effects, all models return only an upper limit on the source size parameter $\rho_\star$. We will compare all models in detail in Sect.~\ref{sec:compare}.




\begin{table*}
\begin{center}
\begin{tabular}{ccccccc}
\hline
 & 1c & 1w & 2c & 2w & 3c & 3w \\
\hline
$d$ & $0.29_{-0.03}^{+0.02}$ & $4.4_{-0.3}^{+0.7}$ & \
$0.227_{-0.008}^{+0.011}$ & $6.3_{-0.3}^{+0.8}$ & \
$0.455_{-0.006}^{+0.005}$ & $2.6_{-0.1}^{+0.2}$  \\
$q$ & $0.14_{-0.04}^{+0.03}$ & $0.19_{-0.05}^{+0.08}$ & \
$0.31_{-0.06}^{+0.05}$ & $0.6_{-0.2}^{+0.3}$ & \
$0.095_{-0.006}^{+0.004}$ & $0.12_{-0.02}^{+0.02}$  \\
$u_0$ & $0.060_{-0.005}^{+0.003}$ & $-0.6_{-0.3}^{+0.2}$ & \
$0.048_{-0.003}^{+0.003}$ & $-0.6_{-0.3}^{+0.2}$ & \
$-0.089_{-0.004}^{+0.006}$ & $-0.21_{-0.03}^{+0.02}$  \\
$\theta$ & $1.945_{-0.007}^{+0.009}$ & $1.951_{-0.009}^{+0.006}$ & \
$0.28_{-0.01}^{+0.01}$ & $0.275_{-0.005}^{+0.012}$ & \
$2.35_{-0.01}^{+0.01}$ & $2.49_{-0.03}^{+0.05}$  \\
$t_0$ & $4688.691_{-0.006}^{+0.007}$ & $4694_{-2}^{+3}$ & \
$4688.685_{-0.007}^{+0.006}$ & $4630_{-30}^{+20}$ & \
$4688.523_{-0.007}^{+0.010}$ & $4692.5_{-0.9}^{+1.1}$  \\
$t_\rmn{E}$ & $20.3_{-0.9}^{+1.4}$ & $23_{-1}^{+2}$ & $24_{-2}^{+1}$ & \
$30_{-2}^{+3}$ & $16.4_{-0.6}^{+0.7}$ & $20.6_{-1.0}^{+0.8}$  \\
$\rho_\star$ & $<0.0036$ & $<0.0028$ & $<0.0019$ & $<0.0017$ & $<0.0031$ \
& $<0.0029$  \\
$g$ (OGLE)& $-0.05_{-0.05}^{+0.08}$ & $-0.03_{-0.07}^{+0.06}$ & \
$0.19_{-0.09}^{+0.06}$ & $0.20_{-0.10}^{+0.03}$ & \
$-0.33_{-0.03}^{+0.04}$ & $-0.16_{-0.05}^{+0.03}$  \\
$\chi^2$ & $568.0$ & $569.0$ & $589.5$ & $590.6$ & $592.9$ & $590.0$  \
\\
\hline
\end{tabular}
\caption{The six static binary lens models explaining the peak
anomaly. $d\,\theta_\rmn{E}$ is the binary angular separation, and $q = M_2/M_1$ the mass ratio. The closest angular approach $u_0\,\theta_\rmn{E}$ between the source and the centre of mass of the binary lens occurs at time $t_0$. The angle $\theta$ measures the 
orientation of the source trajectory from the binary axis, and $t_\mathrm{E}$ is the event time-scale. Finally, $\rho_\star\,\theta_\rmn{E}$ is the angular radius of the source star, and $g$ the blend ratio, i.e. the quotient of background and source flux. $t_\rmn{E}$ is in units of days, $t_0$ is HJD$-$2450000, $\theta$ is in radians, and all other parameters are dimensionless. We have rescaled the photometric uncertainties so that the $\chi^2$ of model 1c matches the number of degrees of freedom (568). The quoted parameter intervals correspond to $\chi^2$ levels
that include 68\,\% of the Markov Chain realisations; for $\rho_\star$ we only give an upper limit at 68\,\% confidence level. While a non-negative OGLE blend ratio is compatible with models 1 and 2, models 3 need adjustment.}\label{Tab:static}
\end{center}
\end{table*}

\begin{table*}
\begin{center}
\begin{tabular}{ccccc}
\hline
& 1c$^+$ & 1w$^+$ & 2c$^+$ & 2w$^+$ \\
\hline
$d$ & $0.27_{-0.03}^{+0.02}$ & $4.6_{-0.1}^{+0.9}$ &
$0.227_{-0.008}^{+0.010}$ & $6.4_{-0.4}^{+0.9}$  \\
$q$ & $0.15_{-0.05}^{+0.04}$ & $0.21_{-0.05}^{+0.07}$ &
$0.31_{-0.07}^{+0.04}$ & $0.61_{-0.22}^{+0.38}$  \\
$u_0$ & $0.0567_{-0.0028}^{+0.0003}$ & $-0.67_{-0.56}^{+0.21}$ &
$0.048_{-0.003}^{+0.003}$ & $-0.62_{-0.36}^{+0.21}$  \\
$\theta$ & $1.948_{-0.006}^{+0.008}$ & $1.951_{-0.006}^{+0.009}$ &
$0.28_{-0.01}^{+0.01}$ & $0.278_{-0.008}^{+0.010}$  \\
$t_0$ & $4688.694_{-0.005}^{+0.008}$ & $4696_{-2}^{+6}$ &
$4688.684_{-0.006}^{+0.007}$ & $4620_{-40}^{+30}$  \\
$t_\rmn{E}$ & $21.4_{-0.2}^{+1.0}$ & $23.85_{-0.06}^{+1.76}$ & $24_{-1}^{+1}$ &
$30_{-2}^{+3}$  \\
$\rho_\star$ & $<0.0039$ & $<0.0033$ & $<0.0022$ & $<0.0018$  \\
$g$ (OGLE) & $<0.047$ & $<0.049$ & $0.18_{-0.08}^{+0.07}$ &
$0.18_{-0.09}^{+0.05}$  \\
$\chi^2$ & $568.6$ & $569.3$ & $589.6$ & $590.8$  \\
\hline
\end{tabular}
\caption{Viable static binary lens models with non-negative OGLE blend ratio imposed. While there is little effect on models 1 and 2, models 3 have dropped out with no corresponding minima found.}
\label{Tab:staticplus}
\end{center}
\end{table*}

\subsection{Binary-source models}

As anticipated, the binary source interpretation might be able to explain the anomaly in OGLE-2008-BLG-510 as efficiently as the binary lens. If we neglect the orbital motion, the microlensing light curve of a "static" binary source with a single lens
is just the superposition of two Paczy\'{n}ski curves with the same $t_\rmn{E}$, so that
\begin{equation}
A(t) = (1-\omega)\,A[u(t;t_1,u_1,t_\rmn{E})] + \omega\,A[u(t;t_2,u_2,t_\rmn{E})]\,,
\end{equation}
where $\omega = F_{\rmn{S},2}/(F_{\rmn{S},1} + F_{\rmn{S},2})$ is the flux offset ratio for the source fluxes
$F_{\rmn{S},1}$ and $F_{\rmn{S},2}$, and we just need to distinguish two pairs $(u_1,t_1)$ and $(u_2,t_2)$ that
characterise the closest angular approach to each of the constituents.

As pointed out by \citet[Appendix C]{Do:MLMC1}, there is a two-fold ambiguity for the angular separation $\Delta \eta$ between the source stars, given that the photometric light curve does not tell us whether they are on the same side (`cis' configuration) or on opposite sides (`trans' configuration) of the source-lens trajectory, where
\begin{equation}
\Delta \eta = \theta_\rmn{E}\,\sqrt{\left(\frac{t_2-t_1}{t_\rmn{E}}\right)^2+\left(u_2 \pm u_1\right)^2}\,.
\label{eq:binsep}
\end{equation}


The search in the parameter space is much simpler, since there is
only one way of superposing two Paczy\'{n}ski curves so as to obtain
the asymmetric peak of OGLE-2008-BLG-510. We found that a small negative blend ratio is preferred, where the parameters shift only very little if a non-negative blend ratio is enforced. The best-fitting model with this constraint is given in
Table \ref{Tab source}. 

\begin{table}
\begin{center}
\begin{tabular}{ccc}
\hline
& bs$^+$ & bs$^+$/$\pi$ \\
\hline
$u_{1}$ & $0.0746_{-0.0026}^{+0.0005}$ & $0.0745_{-0.0033}^{+0.0006}$  \\
$u_{2}$ & $0.0165_{-0.0008}^{+0.0004}$ & $0.0162_{-0.0007}^{+0.0009}$  \\
$t_{1}$ & $4688.45_{-0.02}^{+0.01}$ & $4688.45_{-0.02}^{+0.02}$  \\
$t_{2}$ & $4689.110_{-0.005}^{+0.004}$ & $4689.112_{-0.009}^{+0.006}$  \\
$\omega$ & $0.159_{-0.006}^{+0.010}$ & $0.157_{-0.004}^{+0.011}$  \\
$t_\rmn{E}$ & $21.67_{-0.10}^{+0.69}$ & $22.0_{-0.6}^{+0.6}$  \\
$\pi_{\rmn{E},\parallel}$ & $$ & $-3.6_{-3.1}^{+9.9}$  \\
$\pi_{\rmn{E},\perp}$& $$ & $-0.20_{-0.53}^{+0.42}$  \\
$\pi_\rmn{E}$  & $$ & $<5.1$  \\
$g$ (OGLE)& $<0.030$ & $<0.040$  \\
$\chi^2$ & $582.2$ & $581.7$  \\
\hline
\end{tabular}
\caption{Binary-source model without and with parallax, where we constrain the OGLE blend ratio to be non-negative. The times $t_1$ and $t_2$ are in $\rmn{HJD} - 2\,450\,000$, while all other parameters are dimensionless.} \label{Tab source}
\end{center}
\end{table}

\subsection{Higher order effects}

Beyond the basic static binary-lens and binary-source models presented above, we looked for potential signatures of three possible higher order effects: annual parallax, lens orbital motion and source limb darkening.

As the Earth revolves around the Sun, the trajectory of the source relative to the lens system effectively becomes curved. The curvature
depends on the orientation of the source trajectory relative to the ecliptic, which can be expressed by a parallax vector ${\vec  \pi}_\rmn{E}$ \citep[e.g.\ ][]{Gould:jerk}  with the absolute value 
\begin{equation}
\pi_\rmn{E} = \frac{\pi_\rmn{LS}}{\theta_\rmn{E}} = \sqrt{\pi_{\rmn{E},\parallel}^2+\pi_{\rmn{E},\perp}^2}\,,
\end{equation}
where the components $(\pi_{\rmn{E},\parallel},\pi_{\rmn{E},\perp})$ parallel and perpendicular
to the source trajectory characterise the relative direction of the ecliptic. A typical signature of the annual parallax effect in microlensing light curves is an asymmetry between the rise and the descent.


Starting from the static solutions of Tab. \ref{Tab:staticplus}, we have run Markov chains including the two parallax parameters, where our results are summarised in Tab. \ref{Tab parallax}. For models 1, $\chi^2$ reduces by less than one with the two additional parameters, which shows the insignificance of parallax. There is a larger $\Delta \chi^2$ for models 2, namely 7.3 and 2.5, respectively, but we need to consider that improvements at such level can easily be driven by data systematics (not following the assumption of uncorrelated normally-distributed measurements). We therefore only use an upper limit on the parallax for constraining the physical nature of the lens system. Similarly, we do not find any significant parallax signal with the binary-source model (see Table \ref{Tab source}) and a similarly weak limit.


\begin{table*}
\begin{center}
\begin{tabular}{ccccc}
\hline
& 1c$^+$/$\pi$ & 1w$^+$/$\pi$ & 2c$^+$/$\pi$ & 2w$^+$/$\pi$ \\
\hline
$d$ & $0.28_{-0.04}^{+0.01}$ & $4.73_{-0.36}^{+0.09}$ &
$0.201_{-0.002}^{+0.022}$ & $6.22_{-0.07}^{+0.08}$  \\
$q$ & $0.14_{-0.05}^{+0.05}$ & $0.2217_{-0.0602}^{+-0.0008}$ &
$0.38_{-0.11}^{+0.04}$ & $0.55_{-0.03}^{+0.03}$  \\
$u_0$ & $0.0568_{-0.0045}^{+0.0003}$ & $-0.72_{-0.02}^{+0.23}$ &
$0.041_{-0.003}^{+0.006}$ & $-0.48_{-0.11}^{+-0.02}$  \\
$\theta$ & $1.948_{-0.007}^{+0.008}$ & $1.88_{-0.02}^{+0.06}$ &
$0.306_{-0.029}^{+0.009}$ & $0.24_{--0.01}^{+0.06}$  \\
$t_0$ & $4688.69_{-0.01}^{+0.02}$ & $4694.58_{-1.34}^{+0.05}$ &
$4688.695_{-0.011}^{+0.010}$ & $4625_{-1}^{+2}$  \\
$t_\rmn{E}$ & $21.4_{-0.2}^{+1.6}$ & $23.4_{-0.3}^{+1.1}$ & $28.8_{-4.7}^{+0.8}$ &
$31_{-1}^{+1}$  \\
$\rho_\star$ & $<0.0038$ & $<0.0047$ & $<0.0020$ & $<0.0018$  \\
$\pi_{\rmn{E},\parallel}$ & $-2.1_{-5.1}^{+7.3}$ & $2.4_{-1.8}^{+1.7}$ &
$-8.0_{-1.9}^{+3.0}$ & $0.35_{-0.41}^{+-0.09}$  \\
$\pi_{\rmn{E},\perp}$ & $-0.27_{-0.64}^{+0.40}$ & $-0.60_{-0.31}^{+0.47}$ &
$-1.3_{-0.3}^{+0.7}$ & $0.35_{-0.04}^{+0.14}$  \\
$\pi_\rmn{E}$ & $<5.5$ & $<3.1$ & $<8.0$ & $<0.47$  \\
$g$ (OGLE) & $<0.082$ & $<0.054$ & $0.41_{-0.22}^{+0.07}$ &
$0.17_{-0.06}^{+0.07}$  \\
$\chi^2$ & $567.9$ & $568.3$ & $582.3$ & $588.3$  \\
\hline
\end{tabular}
\caption{Models 1c$^+$, 1w$^+$, 2c$^+$, and 2w$^+$ including the annual parallax effect. The parameters are described in Tab. \ref{Tab:static}, with the addition of the parallax parameters $\pi_{\rmn{E},\parallel}$ and $\pi_{\rmn{E},\perp}$. For the total parallax $\pi_\rmn{E}$ we give the upper limit at $68\%$ confidence level.} \label{Tab parallax}
\end{center}
\end{table*}


The orbital motion of the lens might particularly affect our candidate models in close-binary topology, where a relatively short orbital period compared to the event time-scale $t_\rmn{E}$ is possible and worth being checked for. We have therefore performed an extensive search for candidate orbital configurations, including the three velocity components of the secondary lens in the centre of mass frame as additional parameters. Due to the lack of evident signatures of orbital motion, we have just considered circular orbits, which are completely determined by the three parameters just introduced.
Indeed, we find particular solutions with a sizeable improvement in the $\chi^2$ with respect to all static models. For example, model 1c gets down to $\chi^2=539$. However, a closer look at the candidate solutions thus found shows that they are actually fitting the evident systematics in the data taken during the descent of the microlensing light curve, when the Moon was close to the source star. In all these data, a weak overnight trend is present, whose steepness depends on the particular image reduction pipeline employed. Again, we see that systematics in the data prevent us from drawing firm conclusions just from improvements in $\chi^2$, not knowing whether the data related to it show a genuine signal. Consequently, we will withdraw the hypothesis that orbital motion can be detected, and we will not use any orbital motion information in the interpretation of the event.

We have also looked into source orbital motion, and not that surprisingly, we also find that the fit is driven by systematic trends of the data over the night. Therefore, we will not consider the improvement in the $\chi^2$ down to 544 as due to intrinsic physical effects.

Finally, we have tested several source brightness profiles to account for limb darkening for all models, but the difference with the uniform brightness models is tiny, orders of magnitude smaller than the photometric uncertainty. This is consistent with the fact that in all models the source passes relatively far from the caustic, leaving no hope to study the details of the source structure.

\begin{figure*}
\centering{ \resizebox{8cm}{!}{\includegraphics{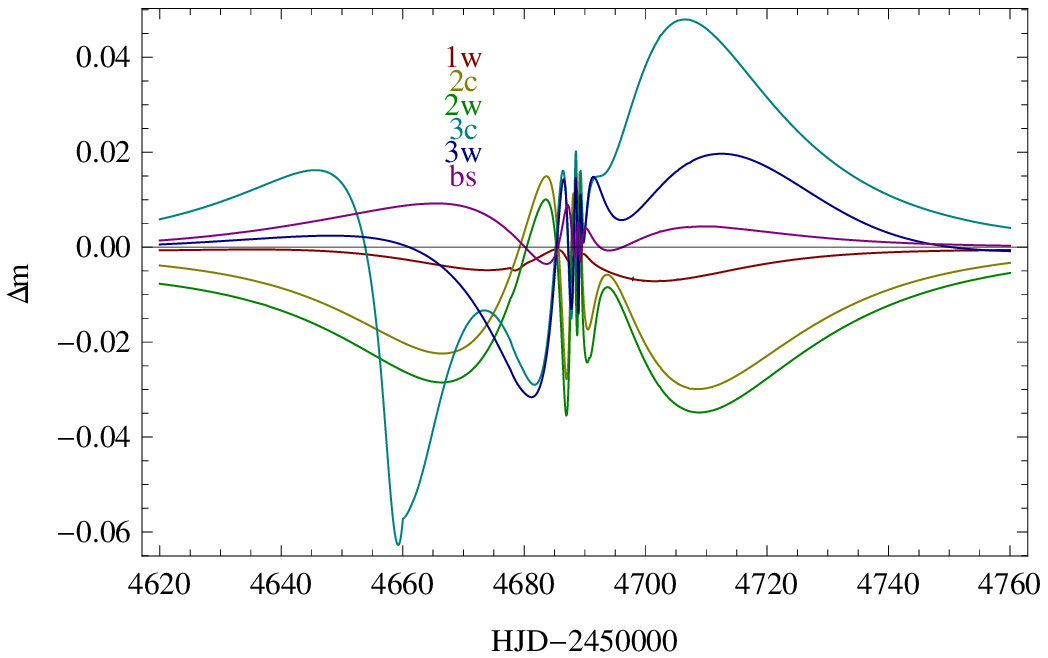}} \resizebox{8cm}{!}{\includegraphics{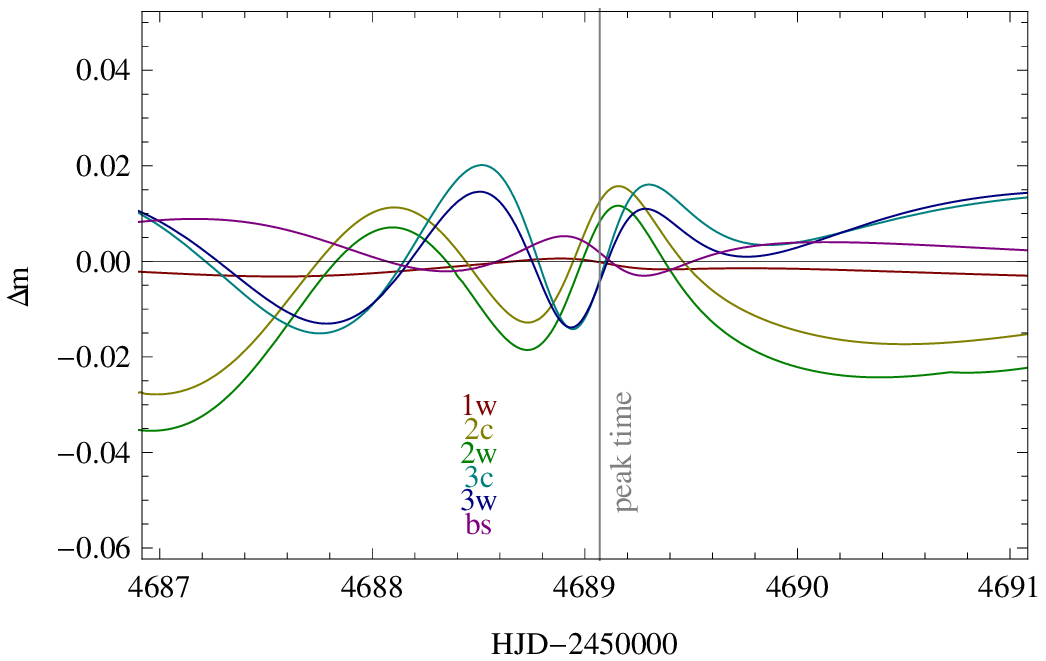}} }
 \caption{Observable magnitude shift $\Delta m = 2.5 \lg \{[A(t)+g]/(1+g)\}$ as a function of time for all suggested models as compared to model 1c. The OGLE blend ratio has been used as reference.}
 \label{Fig:delmagobs}
\end{figure*}

\begin{figure*}
\centering{ \resizebox{14cm}{!}{\epsfig{file=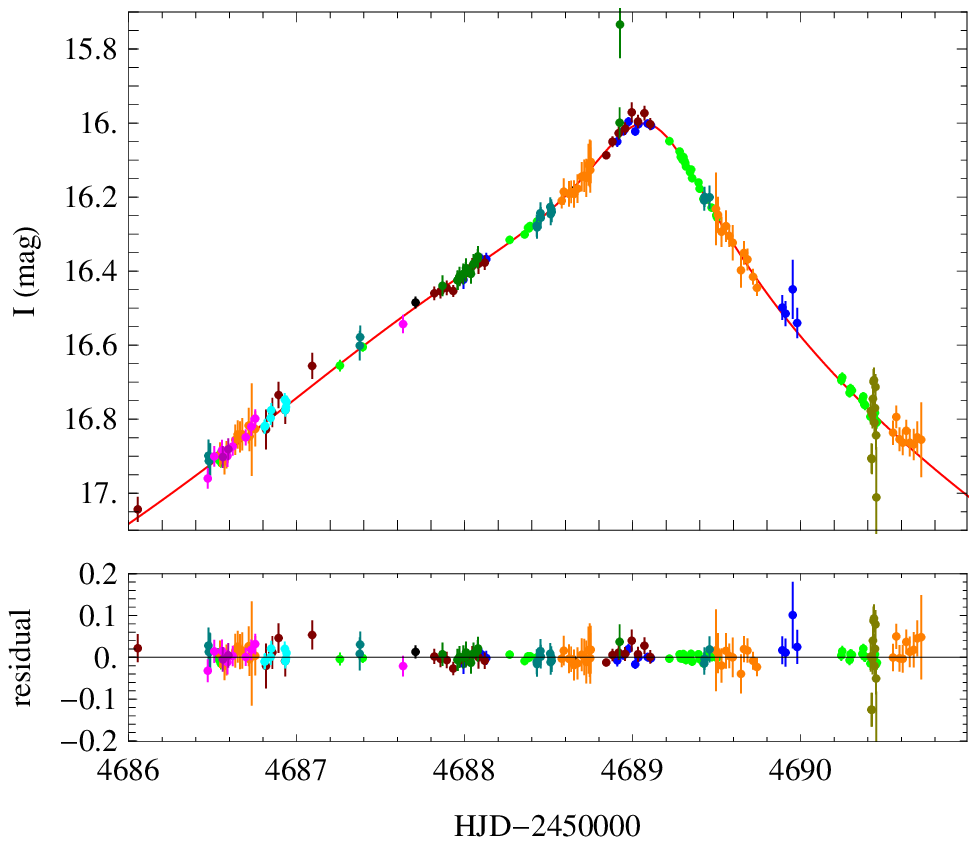,clip=true}} \raisebox{1.78cm}{\Large{1c}}}
\centering{ \resizebox{14cm}{!}{\epsfig{file=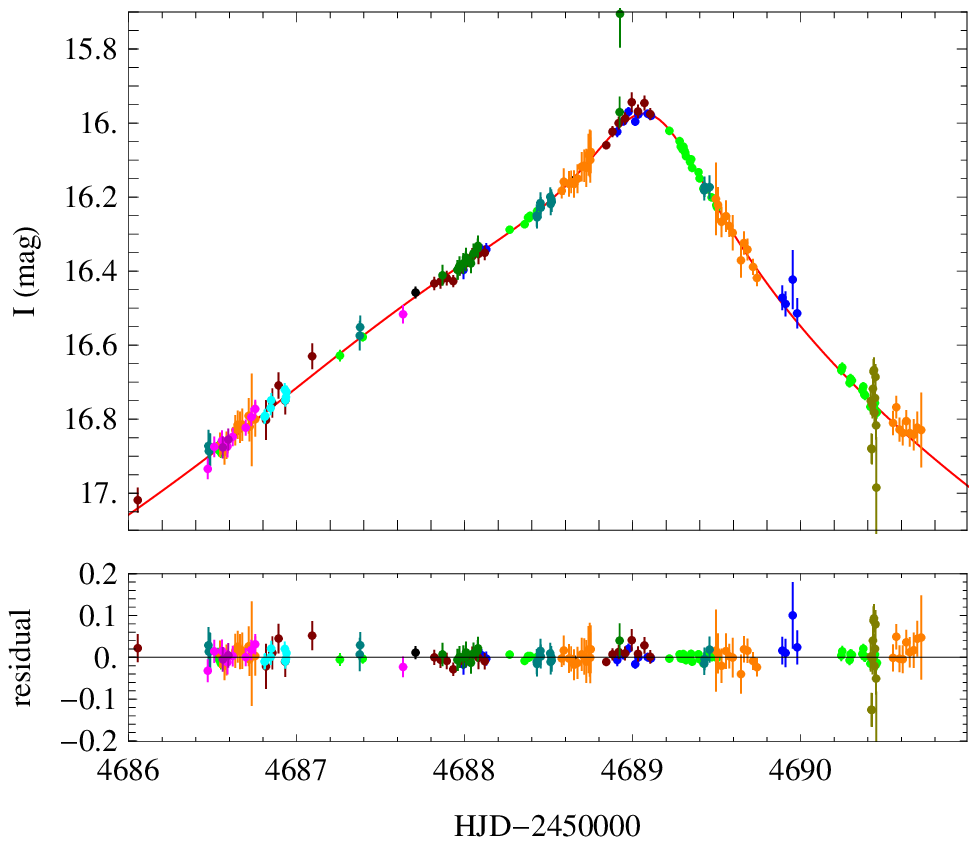,clip=true}} \raisebox{1.78cm}{\Large{1w}}}
\centering{ \resizebox{14cm}{!}{\epsfig{file=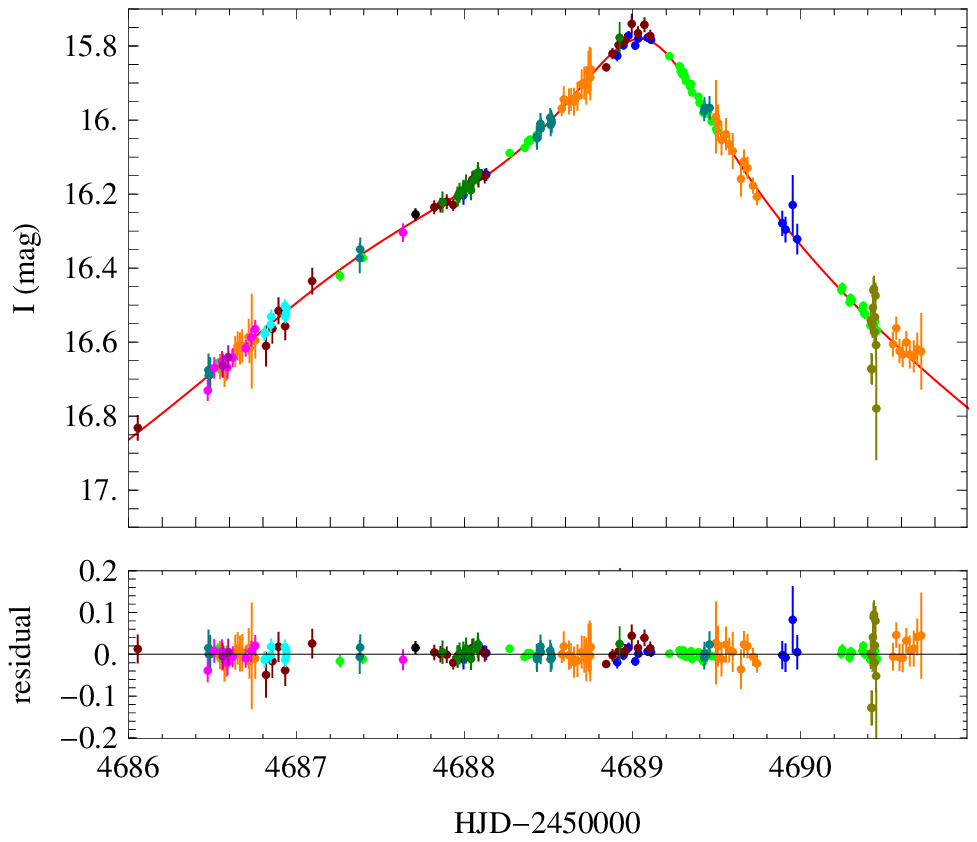,clip=true}} \raisebox{1.78cm}{\Large{2c}}}
\centering{ \resizebox{14cm}{!}{\epsfig{file=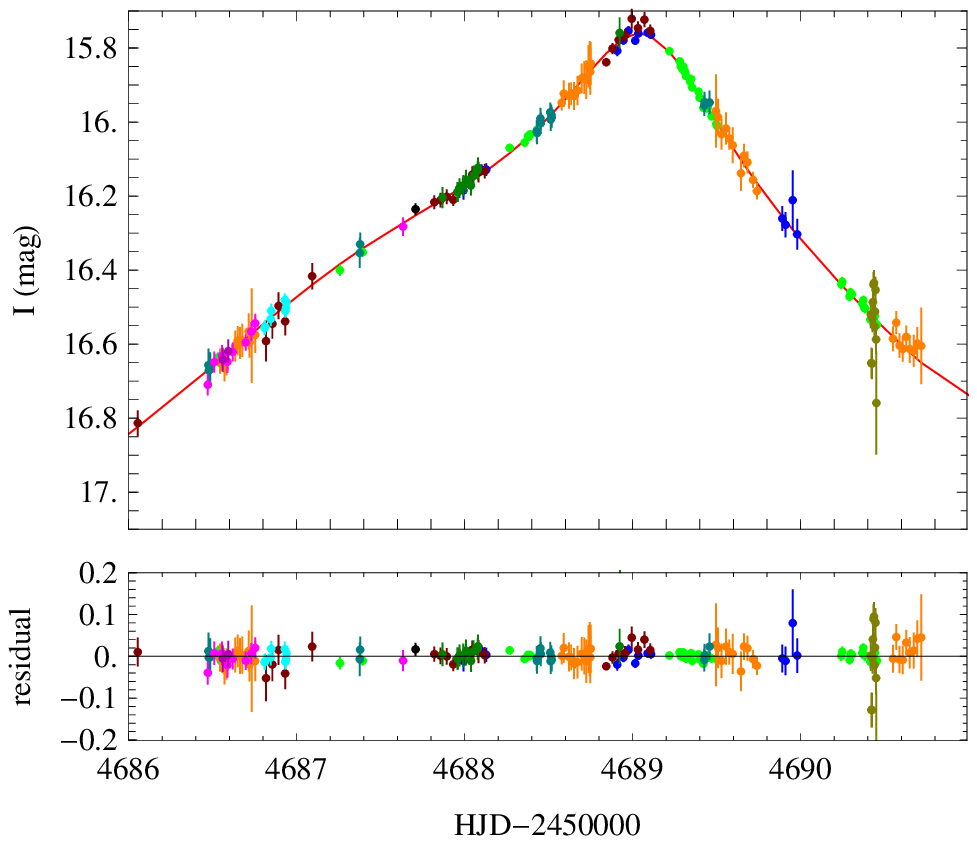,clip=true}} \raisebox{1.78cm}{\Large{2w}}}
\centering{ \resizebox{14cm}{!}{\epsfig{file=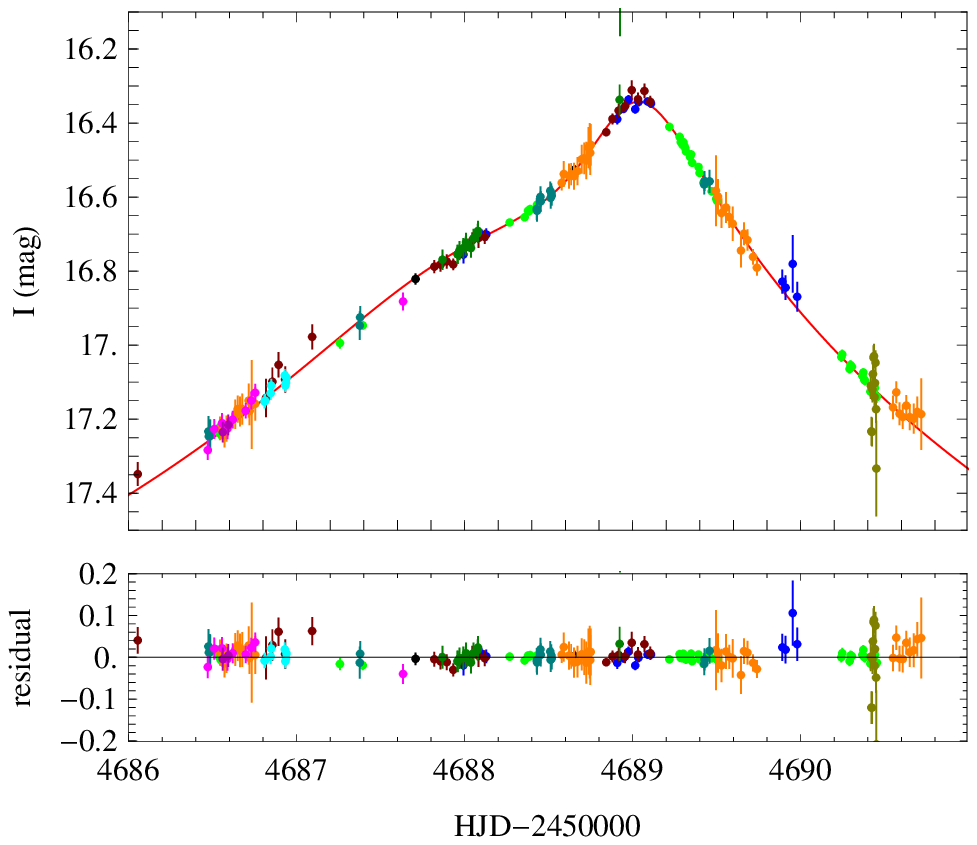,clip=true}} \raisebox{1.78cm}{\Large{3c}}}
\centering{ \resizebox{14cm}{!}{\epsfig{file=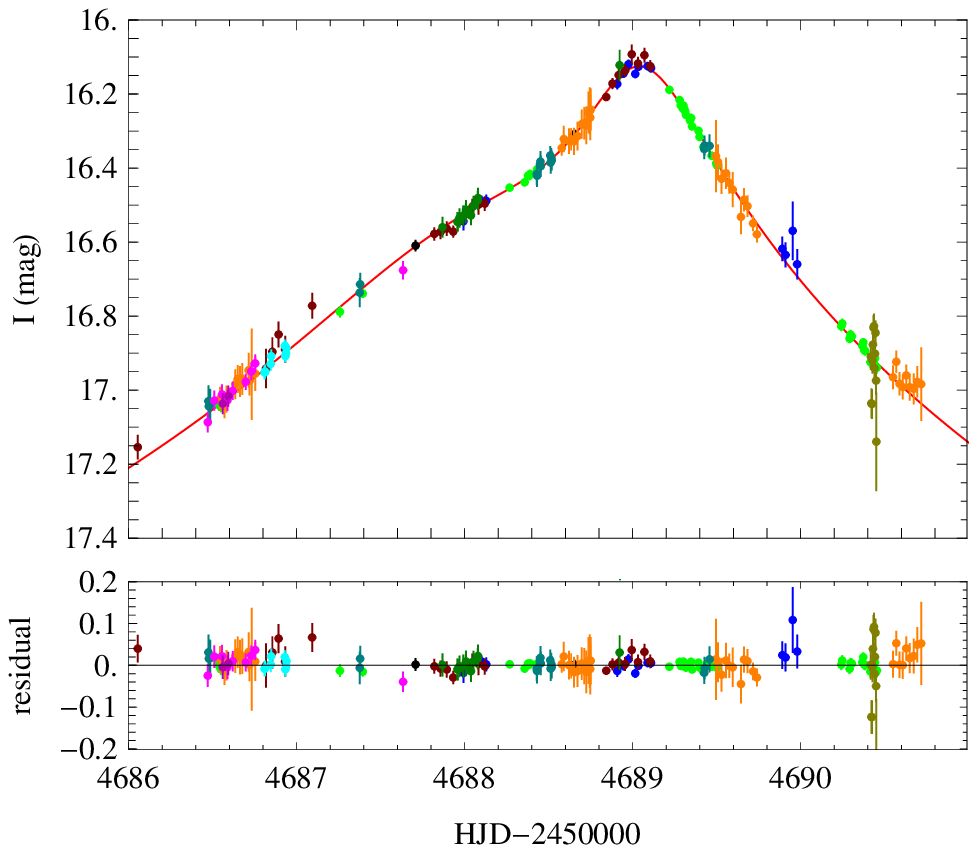,clip=true}} \raisebox{1.78cm}{\Large{3w}}}
\centering{ \resizebox{14cm}{!}{\epsfig{file=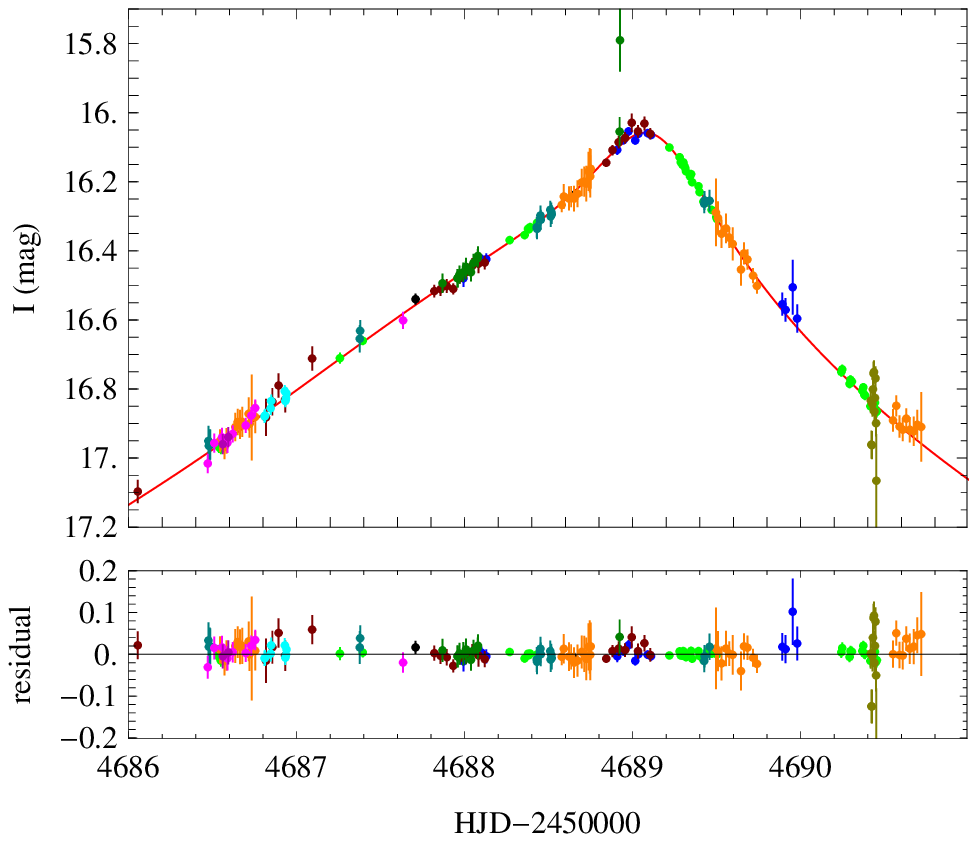,clip=true}} \raisebox{1.78cm}{\Large{bs}}}
 \caption{Residuals in the peak region of OGLE-2008-BLG-510 for all proposed models.}
 \label{Fig:residcompare}
\end{figure*}

\subsection{Limitations due to systematics in the data}

The photometric analysis for this event posed several challenges. The
source star was very faint ($I \sim 19.23$) and heavily blended with a
brighter companion. A saturated star was nearby, which was necessarily
masked by our photometric pipeline, but which in turn placed
constrains on the size of the PSF and the kernel that could be used
during the image subtraction stage, particularly for the images that
had high seeing values. In a few specific cases, using a larger fit
radius would mean that the masked area around the saturated star fell
within the fit radius of the source star, thereby reducing the number of
pixels used in the photometry \citep{Bramich2011}.
A new version that is currently under
development discounts a bigger region of pixels
around the saturated star from the kernel solution, since the
kernel solution is most sensitive to contaminated pixels.

Shortly after the ongoing anomaly had been identified, the moon was 
full and close to the target field of observation resulting in high sky
background counts and a strong background gradient. While the
pipeline can account for the latter, the former has an impact on the
photometric accuracy that can be achieved. In fact, 
 the moon started to have a strong effect from 11 August 2008
(illumination fraction 80\%) and $\sim\,$7--10$^\circ$ from the target field,
on to 12 August 2008 (illumination fraction 86\%) and $\sim\,$3--7$^\circ$ from the
target field. It was full on 16 August 2008. So all the affected data were
taken after the anomaly at the peak. Normally it is not advisable to observe if the moon is bright and
closer than 15$^\circ$ to the target. Obtaining a full characterisation of the impact of Moon pollution
on the photometry is not
that straightforward to assess with difference imaging, because there are many other factors 
that affect the kernel solution as well, and the different contributions are not
easy to isolate. We have however optimised our photometry to minimise the impact of
Moon pollution and accounted for the uncertainty during our modelling
runs. Moreover, {\sc DanDIA} takes care of systematic effects by producing a $\chi^2$ value for the star fits and
reflecting this in the reported size of the photometric error bars.

\subsection{Comparison of models}
\label{sec:compare}



For the peak region, the differences between the models remain below 2.5\,\% (see Fig.~\ref{Fig:delmagobs}), which makes these difficult to probe with the apparent scatter of the data. Moreover, only if it is ensured that systematic effects are consistently substantially below this level, will a meaningful discrimination between the competing models be possible. In our opinion, this poses the most substantial limit to our analysis, and we therefore abstain from drawing very definite conclusions about the physical nature of the OGLE-2008-BLG-510 microlens. Fig.~\ref{Fig:delmagobs} also shows that the differences between the respective pairs of close- and wide-binary models are even much smaller, less than 0.4 to 0.7\,\% over the peak. It can also be seen that models are made to coincide better in the peak region where more data have been acquired, as compared to the wings where larger differences of 3--6\,\% are tolerated.

As Fig.~\ref{Fig:residcompare} shows, the differences between the models with respect to the residuals with observed data appear to be rather subtle. Additional freedom that causes further narrowing arises from adopting a blend ratio for each of the sites and each of the models, allowing for different relations between the blend ratios between the models. A successful fit should be characterised by data falling randomly to both sides of the model light curve. On this aspect, models 2 appear better balanced between all data sets than models 1, in particular with respect to the Canopus 1.0m and the MOA data, but it introduces a trend in some SAAO data over the peak that does not show with other models. We also see that models 3 are more closely related to models 2 than to models 1, with rather similar behaviour over the peak, while the differences in the wing regions are related to the different blend ratios. In fact, both models 2 and 3 have the second and closer cusp approach to the off-axis cusp, whereas the roles of the on-axis and off-axis cusps are flipped for models 1.

\begin{figure}
\centering{ \resizebox{8cm}{!}{\includegraphics{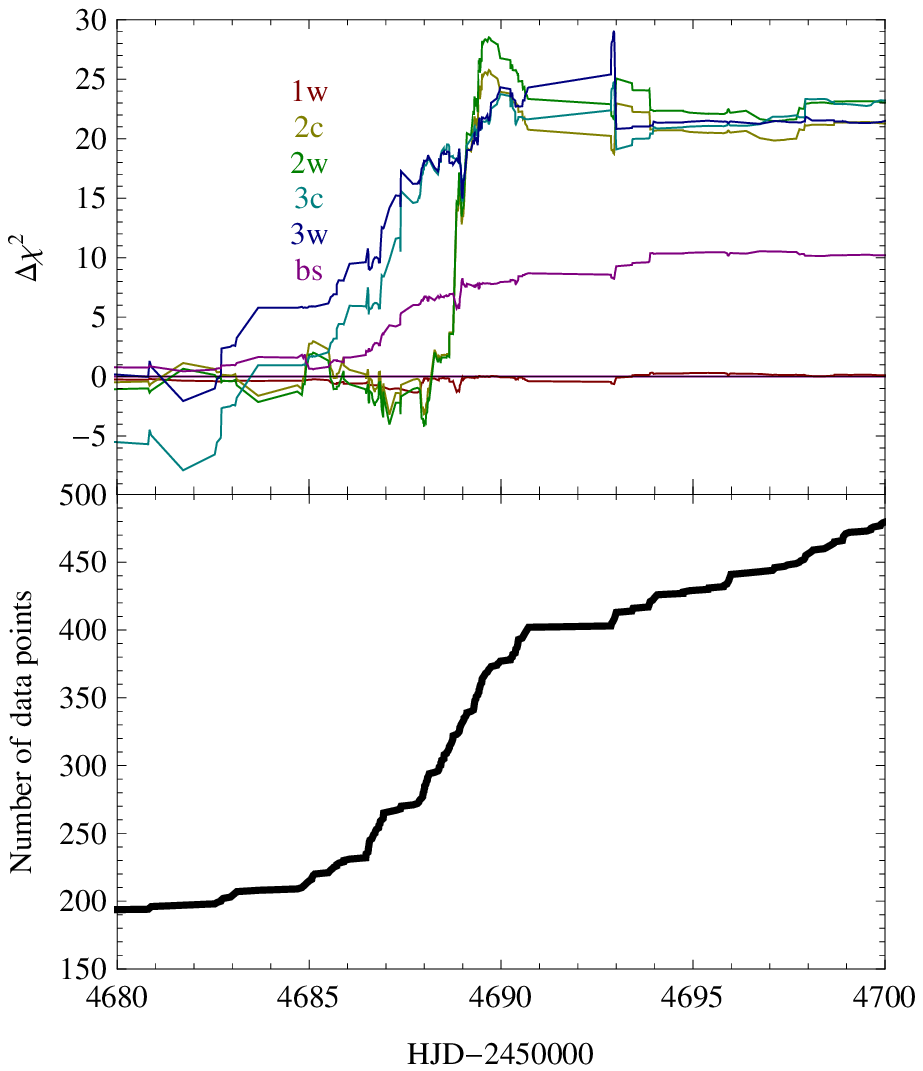}} }
 \caption{$\Delta \chi^2$ as a function of time for all suggested models relative to model 1c, along with the cumulative number of data points up to that time.}
 \label{Fig:delchisq}
\end{figure}

It is also very instructive to look at $\Delta \chi^2$ between the models as a function of time as the data were acquired (see Fig.~\ref{Fig:delchisq}), which obviously depends on the sampling. While models 3 as well as the binary-source model rather gradually lose out to models 1 for $4683 \leq \rmn{HJD}-2450000 \leq 4691$, models 2 perform much worse than all other models for the rather short period $4688.0 \leq \rmn{HJD}-2450000 \leq 4689.5$, which happens just to coincide with the skewed peak. However, for  $4686.0 \leq \rmn{HJD}-2450000 \leq 4688.0$ and $4689.5 \leq \rmn{HJD}-2450000 \leq 4690.5$, models 2 do a better job than models 1. It is in the nature of $\chi^2$ minimisation that more weight is given to regions with a denser coverage by data. As a consequence, relative $\chi^2$ between competing regions of parameter space depend on the specific sampling. In fact, the region $4689.5 \leq \rmn{HJD}-2450000 \leq 4690.5$ which favours model 2 has a sparser coverage than preceding nights. Moreover, we find model 3c outperforming all other models for $4645 \leq \rmn{HJD}-2450000 \leq 4682$, but this is given little weight. The weight arising by the sampling is a particularly relevant issue given that for non-linear models, the leverage to a specific parameter strongly depends on the time observations are taken. We immediately see that there is an extreme danger of conclusions being driven by systematics in a single data set during a single night, potentially overruling all that we learn from other data.

\section{Physical interpretation}



An ordinary microlensing event is determined by the fluxes of the source star $F_\rmn{S}$, the blend $F_\rmn{B}$, the mass of the lens star $M$, the source and lens distances $D_\rmn{S}$ and $D_\rmn{L}$, as well as the relative proper motion $\mu$ between lens and source star (or the corresponding effective lens velocity $v = D_\rmn{L}\,\mu$), and finally the angular source-lens impact $\theta_0$ and its corresponding epoch $t_0$. However, the photometric light curve is already fully described by a smaller number of parameters. Apart from $F_\rmn{S}$, and $F_\rmn{B}$, it is completely characterised by $t_0$, $u_0 = \theta_0/\theta_\rmn{E}$ and $t_\rmn{E} = \theta_\rmn{E}/\mu$, where the angular Einstein radius $\theta_\rmn{E}$, as given by Eq.~(\ref{eq:thetaE}) absorbs $M$, $D_\rmn{L}$, and $D_\rmn{S}$. All the physics of $D_\rmn{S}$, $D_\rmn{L}$, $M$, and $v$ gets combined into the single model parameter $t_\rmn{E}$, and information about the detailed physical nature gets lost.

Higher-order effects can play an important role in recovering the information. Additional relations between the physical properties can be established if the light curve depends on further parameters that are related to the angular Einstein radius $\theta_\rmn{E}$ or the relative lens-source parallax $\pi_\rmn{LS}$. In particular, finite-source effects with the parameter $\rho_\star = \theta_\star/\theta_\rmn{E}$ or annual parallax with the microlensing parallax $\pi_\rmn{E} = \pi_\rmn{LS}/\theta_\rmn{E}$ allow solving for $D_\rmn{L}$, $M$, and $v$, once $D_\rmn{S}$ and the angular size $\theta_*$ of the source star can be established. Moreover, as in the case of OGLE-2008-BLG-510 here, even the absence of detectable effects can provide valuable constraints to parameter space. In fact, as discussed in the previous section, the anomaly of OGLE-2008-BLG-510 is almost indifferent to the source radius. Moreover, no parallax signal is clearly detected in either model. Finally, our attempts to find solutions with orbital motion just bumped into systematic overnight trends in the data. 


But even if the physical properties of the lens system cannot be determined directly, Bayes' theorem provides a means of deriving their probabilty density. In fact, with the physical properties $\vec \psi$ and the model parameters $\vec p$, one finds a probability density
$P_{\vec{\psi}}(\vec{\psi}|\vec{p})$ of $\vec \psi$ given $\vec p$ as
\begin{equation}
P_{\vec \psi}(\vec{\psi}|\vec{p})=\frac{{\cal L}(\vec{p}|\vec{\psi})\,P_{\vec{\psi}}(\vec{\psi})}{\int {\cal L}(\vec{p}|\vec{\psi'})\,P_{\vec{\psi'}}(\vec{\psi'})\,\rmn{d}{\vec{\psi'}}}\,,
\label{eq:bayes}
\end{equation}
where $P_{\vec{\psi}}(\vec{\psi})$ is the prior probability density of the physical properties $\vec \psi$, and 
${\cal L}(\vec{p}|\vec{\psi})$ is the likelihood for the parameters $\vec p$ to arise from the properties $\vec \psi$.

While prior probability densities  $P_{\vec{\psi}}(\vec{\psi})$ for the physical properties are straightforwardly given by a kinematic model of the Milky Way and mass function of the various stellar populations, the likelihood ${\cal L}(\vec{p}|\vec{\psi})$ is proportional to the differential microlensing rate $\rmn{d}^k\Gamma/(\mathrm{d}p_1 \ldots \mathrm{d}p_k)$ \citep[c.f.\ ][]{Dominik06}, which is to be evaluated using the constraining relations between the model parameters $\vec p$ and the physical properties $\psi$. We determine $P_{\vec{\psi}}(\vec{\psi})$ following the lines of \citet{Dominik06}. While \citet{Calchi} have recently discussed Galactic models in some detail, we basically follow the choice of  \citet{Grenacher99}. In particular, we consider candidate lenses in an exponential disk or in a bar-shaped bulge. The respective mass functions are taken from \citet{Chabrier03}. The only difference with respect to \citet{Dominik06} is that we consider an anisotropic velocity dispersion for stars in the bulge \citep{HanGould95}, with $\vec{\sigma}=(116,90,79)$ km/s along the three bar axes $\hat X',\hat Y',\hat Z'$ defined therein. Actually, this difference does not have any major effects on the final result.
The denominator in Eq.~(\ref{eq:bayes}) just reflects that the probability density is properly normalised, i.e.\ $\int P_{\vec \psi}(\vec{\psi'}|\vec{p})\,\rmn{d}\psi' = 1$, which can be achieved trivially.

In addition to the information contained in the event time-scale $t_\rmn{E}$, we also take into account the constraints arising from the upper limits on the source size, the parallax, and the luminosity of the lens star, as given by the blend ratio $g$.

The angular source radius $\theta_\star$ can be measured using the technique of \citet{Yoo}. The dereddened color and magnitude of
the source are measured from an instrumental colour-magnitude diagram as $[(V-I),I]_{0,\rmn{S}} = (0.63,17.78)$, where we assumed
that the clump centroid is at $[(V-I)_0,M_I]_\rmn{clump} = (1.06,-0.10)$
\citep[][; D.~Nataf, private communication]{Bensby}, that the Galactocentric distance is $R_0=8.0~\rmn{kpc}$ and that the clump centroid
lies $0.2~\rmn{mag}$ closer to us than the Galactic centre, since the field is at Galactic latitude $l=+5.2^\circ$. Therefore, we also set $D_\mathrm{S} = 7.3~\mbox{kpc}$. We then
convert from $V$/$I$ to $V$/$K$ using the colour-colour relations of \citet{BB88}.  Finally, we use the $V$/$K$ colour/surface-brightness relation of \citet{Kervella} to obtain $\theta_\star = 0.80~\umu{}\rmn{as}$.



Rather than just basing our analysis on the best-fitting values, we also take into account the model parameter uncertainties.
From our Markov-Chain Monte Carlo runs, we find that $t_\rmn{E}$, $\rho_\star$, $\pi_\rmn{E}$ and $g$ (OGLE) are basically uncorrelated.
Their combined probability can therefore fairly be approximated by the product of the individual probability densities 
\begin{equation}
P_{t_\rmn{E},\rho_\star,\pi_\rmn{E},g}
(t_\rmn{E},\rho_\star,\pi_\rmn{E},g)=
P_{t_\rmn{E}}(t_\rmn{E}) P_{\rho_\star}(\rho_\star) P_{\pi_\rmn{E}}(\pi_\rmn{E})P_g(g)\,.
\end{equation}
We approximate each of these distributions by a Gaussian
\begin{equation}
G(x,\bar x,\sigma)=\frac{1}{\sigma\sqrt{2\pi}} e^{-\frac{(x-\bar x)^2}{2\sigma^2}}\,,
\end{equation}
where
\begin{eqnarray}
&& P_{t_\rmn{E}}(t_\rmn{E})=G(t_\rmn{E},\bar t_\rmn{E},\sigma_{t_\rmn{E}}) \nonumber \\
&& P_{\rho_\star}(\rho_\star)= 2\,\Theta(\rho_\star)\,G(\rho_\star,0,\sigma_{\rho_\star}) \nonumber \\
&& P_{\pi_\rmn{E}}(\pi_\rmn{E})=2\,\Theta(\pi_\rmn{E})\,G(\pi_\rmn{E},0,\sigma_{\pi_\rmn{E}})  \nonumber  \\
&& P_g(g)=2\,\Theta(g)\,G(g,0,\sigma_{g})\,,
\end{eqnarray}
with the  dispersions of these Gaussians being identified with the $68\%$ confidence limits listed in Table \ref{Tab parallax}.

\begin{figure*}
\centering{\resizebox{16cm}{!}{\includegraphics{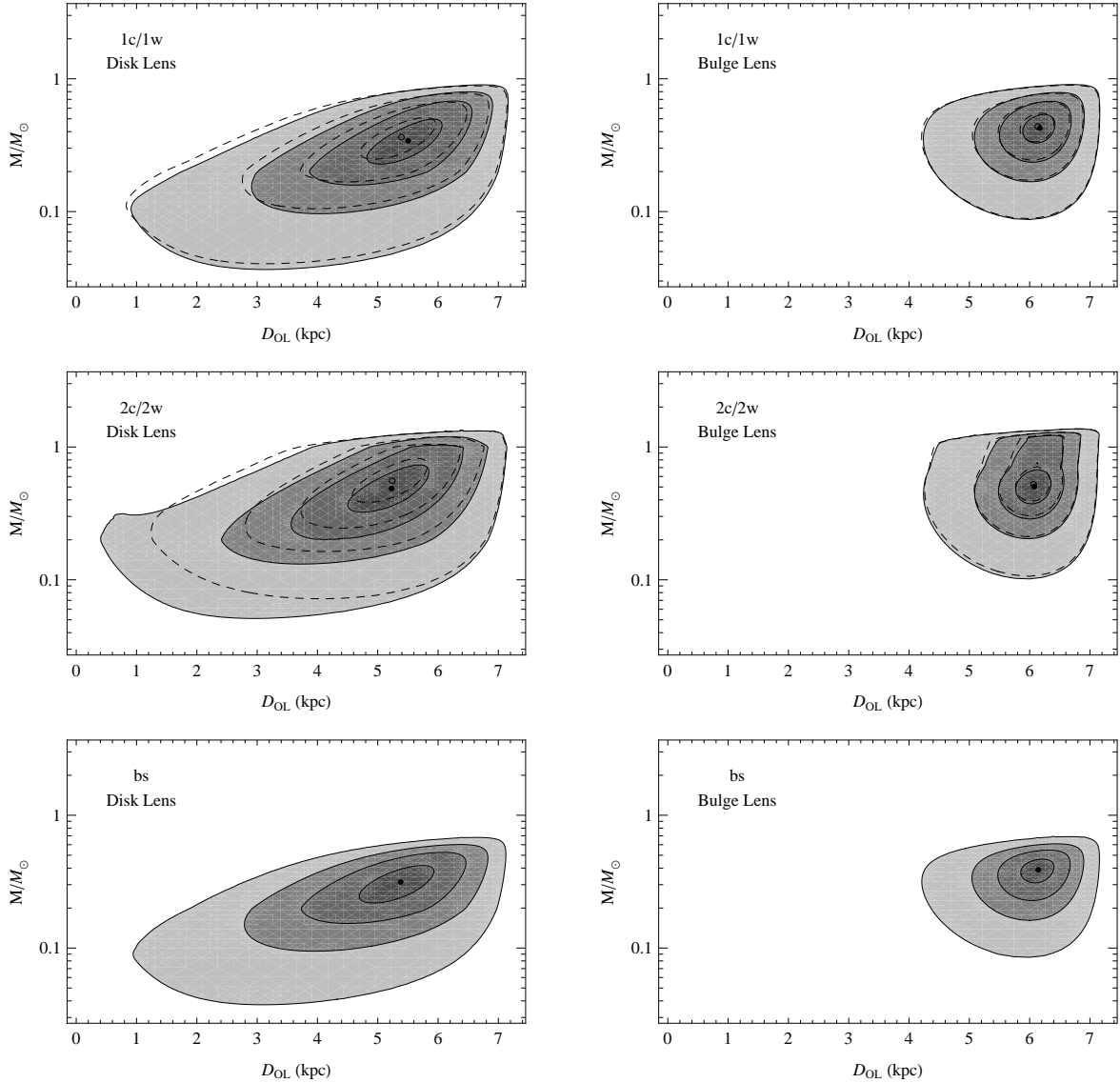}}}
 \caption{Contours of the probability density containing $95\%$, $68\%$, $38\%$, $10\%$ probability in the mass-distance plane for the lens in binary-lens models 1 (top), binary-lens models 2 (middle), and the binary-source model (bottom). The panels on the left treat the case of a disk lens, while those on the right consider a bulge lens. The dashed lines are the contours for models 1w and 2w, respectively. The filled (empty) circles are the modes of the probability density for the close- (wide-)binary models.}
 \label{Fig MDOL12}
\end{figure*}

We find a probability density in $D_\rmn{L}$, $M$, and $v$. By integrating over  $v$, and changing from $M$ to $\log M$, we obtain a probability density in the $(D_\rmn{L}, \log M)$ plane. In this integration, as explained in \citet{Dominik06}, the dispersion in the Einstein time is practically irrelevant and we can approximate the $t_\rmn{E}$ distribution by a delta function.
The arising posterior probabilities are illustrated in Fig. \ref{Fig MDOL12} separately for disk and bulge lenses for models 1 and 2. We can appreciate the small difference between the close- and wide-binary models (whose contours are dashed in Fig. \ref{Fig MDOL12}). The wide topology slightly favours a smaller distance and a higher mass for the lens system. Higher masses are cut off by the blending constraint, as the lens would become too bright to be compatible with the absence of blending. In fact, this blending constraint readily dismisses models 3. The source radius constraint slightly cuts small lens masses as these would yield a too small Einstein radius. Finally, only the tail of the disk distribution at small $D_\rmn{L}$ is affected by the parallax constraint.

The relative weight of disk and bulge for each microlensing model is obtained by integrating these distributions over the whole $(D_\rmn{L},\log M)$ plane. Table \ref{Tab post} shows the probabilities for each microlensing model and each hypothesis for the lens population. We also show the average lens distance and mass in each case. We find that the cut on higher masses due to the absence of blending significantly penalises the bulge, which is known to have a higher microlensing rate thanks to massive lenses. This fact witnesses how the information coming from the study of our specific microlensing event effectively selects the viable lens models.


\begin{table}
\begin{center}
\begin{tabular}{cccccc}
\hline
 Model & Prob. & $\langle D_\rmn{L}\rangle$ & $\langle M\rangle$ & $\langle M_1 \rangle$ & $\langle M_2 \rangle$ \\
            &          & [kpc] & $[M_\odot$] & $[M_\odot]$ & $[M_\rmn{jup}$] \\
\hline
1c Disk & 46.$\%$ & 4.74 & 0.21 & 0.19 & 28\\
1c Bulge & 54.$\%$ & 5.92 & 0.34 & 0.29 & 44\\
\hline
1w Disk & 49.2$\%$ & 4.65 & 0.22 & 0.18 & 42\\
1w Bulge & 50.8$\%$ & 5.9 & 0.34 & 0.28 & 64\\
\hline
2c Disk & 53.8$\%$ & 4.49 & 0.3 & 0.22 & 88\\
2c Bulge & 46.2$\%$ & 5.88 & 0.46 & 0.33 & 133\\
\hline
2w Disk & 53.3$\%$ & 4.66 & 0.36 & 0.23 & 134\\
2w Bulge & 46.7$\%$ & 5.88 & 0.48 & 0.31 & 178\\
\hline
bs Disk & 49.2$\%$ & 4.63 & 0.19 &  & \\
bs Bulge & 50.8$\%$ & 5.9 & 0.29 &  & \\
\hline
\end{tabular}
\caption{Expectation values of lens distance $D_\rmn{L}$ and lens mass $M$ for the different binary-lens models, and assessment of probability for the lens to reside in the Galactic disk or bulge.} \label{Tab post}
\end{center}
\end{table}

Within the parameter uncertainties of the binary-lens models, there is some significant overlap in the possible physical nature of the lens system (with model 2w being compatible with a very wide range). We find that an M dwarf orbited by a brown dwarf appears to be a favoured interpretation, but we cannot fully exclude a more massive secondary. Regardless of the specific model, a location in the Galactic disk appears to be as plausible as in the Galactic bulge.
The microlensing light curve does not definitely clarify whether the system is in a wide or in a close configuration because of the well-known degeneracy of the central caustic. It is interesting to note that if we use the average values for the total mass and the distance we can estimate that the minimum value for the orbital period in model 1c (1w) is 296 days (51 years) with a minimum orbital radius of 0.52 (8.25) AU. Even in the closer configuration, detecting any orbital motion signal in this event would have been quite difficult. This also reinforces our rejection of all models with very short orbital periods found during the modelling phase.

For the binary-source model, we find with the $I$ band magnitude and the luminosity offset ratio $\omega$, the masses of the two components as $m_1 = 1.03~M_{\sun}$ and $m_2 = 0.80~M_{\sun}$. The $V-I$ index of this system is $0.1~\rmn{mag}$ larger than the $V-I$  index of a single source that would yield all the observed flux. Moreover, by adopting values for the lens mass and distance in the middle of the two averages for the bulge or disk hypothesis (see Table~\ref{Tab post}), the physical size of the angular Einstein radius $\theta_\rmn{E}$ at the source distance $D_\rmn{S}$ becomes $D_\rmn{S}\,\theta_\rmn{E} = 2.35~\mbox{AU}$. 
For a face-on circular orbit, we therefore find the orbital radius simply as $r_0 = D_\mathrm{S} \Delta \eta$, where $\Delta \eta$ as given by Eq.~(\ref{eq:binsep}) denotes the angular separation between the binary-source constituents.
With the total mass of the binary-source system $m = m_1 + m_2 \sim 1.83~M_{\sun}$, we then obtain orbital periods of
$P_{0,\rmn{cis}} \sim 16~\mbox{days}$ or $P_{0,\rmn{trans}} \sim 29~\mbox{days}$ for the cis- or trans-configurations, respectively.
For an orbital inclination $i$, the orbital radius increases as $r= r_0/(\cos i)$, so that the orbital periods increase as $P = P_0/(\cos i)^{3/2}$.
For orbits with an eccentricity $\varepsilon$, the observed separation $r$ is limited to the range $a(1-\varepsilon) \leq r \leq a(1+\varepsilon)$, where $a$ denotes the semi-major axis. While $r = a$ for a circular orbit, one finds in particular $a \geq r/2$. Therefore, in the general case, orbital periods still have to fulfil $P \geq P_0/(2\,\cos i)^{3/2}$. Given both the small colour difference and a plausibly long orbital period (as compared to the event time-scale of $t_\rmn{E} \sim 22~\mbox{days}$), there is no apparent inconsistency within the proposed binary-source model.

\section{Conclusions}

So far, most of the results arising from gravitational microlensing campaigns have been based on the detailed discussion of individual events that show prominent characteristic features that allow unambiguous conclusions on the  underlying physical nature of the lens system that caused the event. However, with its sensitivity to mass rather than light, gravitational microlensing is particularly suited to determine population statistics of objects that are undetectable by any other means.This not only includes planets in regions inaccessible to other detection techniques, but also brown dwarfs, low-mass stellar binaries, and stellar-mass black holes. Meaningful population statistics will only arise from controlled experiments with well-defined deterministic procedures, and similarly the data analysis needs to adopt a homogeneous scheme. Given that the information arises from the constraints posed by the data on parameter space, i.e.\ what can be excluded rather than what can be detected, events with weak and potentially ambiguous features need to be dealt with appropriately. 

The event OGLE-2008-BLG-510, which was subject to a detailed analysis in this paper, is a rather typical representative of the large class of events with evident weak features. Our light curve morphology classification approach revealed that the apparent asymmetric shape of the peak appears to be compatible with a range of binary-lens models or a binary-source model. A particular challenge in the interpretation of the acquired data is dealing with systematic effects that can lead to the mis-estimation of measurement uncertainties, non-Gaussianity, and correlations. This in turn strongly limits the validity of the application of statistical techniques that assume uncorrelated and normally (Gaussian) distributed data. We moreover find that the sampling of microlensing light curve affects the preference of regions of parameter space, and that conclusions therefore are at risk to be driven by systematics in the data rather than real effects. In the specific case of OGLE-2008-BLG-510, all suggested models coincide within 2.5\,\% in the peak region (where the sampling is dense), and in order to discriminate, not only does the scatter of the data need to be small, but more importantly, systematics need to be understood or properly modelled to a level substantially below the model differences. For OGLE-2008-BLG-510, we found that systematic trends in some datasets simulate a fake scale for orbital motion, which (if present) is hidden below systematics and thus impossible to determine. 

Despite the fact that the viable binary-lens models differ in the geometry between the binary-lens system and the source trajectory, there is some overlap in the resulting physical nature of the binary-lens system, taking into account the model parameter uncertainties. In fact, we find the lens system consisting of an M dwarf orbited by a brown dwarf a favourable interpretation, albeit that we cannot exclude a different nature. 

The large number of microlensing events with weak features are currently waiting for a comprehensive statistical analysis, and lots of interesting results are likely to be buried there. It however requires a very careful analysis of the noise effects in the data as well as modelling techniques that map the whole parameter space rather than just finding a single optimum, which is just a point estimate, and moreover biased if $\chi^2$ minimisation is used.\footnote{This already holds if all data are uncorrelated and uniformly distributed. Any deviation from these assumptions makes it even worse.} Currently, we neither understand the paucity of reported events due to a binary source nor the impact of brown dwarfs on microlensing events, and we might even need to look into weak signatures and their statistics in order to arrive at a consistent picture. 

The detection of the weak microlensing anomaly in OGLE-2008-BLG-510 also demonstrated the power and feasibility of automated anomaly detection by means of immediate feedback, i.e.\ requesting further data from a telescope which is automatically reduced and photometric measurements being made available within minutes. This is an important step in the efficient operation of a fully-deterministic campaign that gets around the involvement of humans in the decision chain and offers the possibility to derive meaningful population statistics by means of simulations.




\section*{Acknowledgments}

The Danish 1.54m telescope is operated based on a grant from the Danish Natural Science Foundation (FNU).
The ``Dark Cosmology Centre'' is funded by the Danish National Research
Foundation. Work by C. Han was supported by a grant of National Research Foundation of Korea (2009-0081561). L. M. acknowledges support for this work by research funds of the
International Institute for Advanced Scientific Studies. Work by AG was supported by NSF grant AST-0757888.
Work by BSG, AG, and RWP was supported by NASA grant NNX08AF40G. Work by SD was performed under contract with the California Institute
of Technology (Caltech) funded by NASA through the Sagan Fellowship
Program. The MOA team acknowledges support
 by grants JSPS20340052, JSPS20740104 and MEXT19015005. Some of the observations reported in this paper were obtained with the
Southern African Large Telescope (SALT). LM acknowledges support for this work by research funds of the
International Institute for Advanced Scientific Studies. MH acknowledges support by the German Research Foundation (DFG).
DR (boursier FRIA) and JSurdej acknowledge support from the Communaut\'e
fran\c{c}aise de Belgique -- Actions de recherche concert\'ees --
Acad\'emie universitaire Wallonie-Europe. The OGLE project has received funding from the European Research
Council under the European Community's Seventh Framework Programme
(FP7/2007-2013) / ERC grant agreement no. 246678.
MD, YT, DMB, CL, MH, RAS, KH, and CS are thankful to Qatar National Research Fund (QNRF), member of Qatar Foundation, for support by grant NPRP 09-476-1-078.
 
\bibliographystyle{mn2e}
\bibliography{OB08510}









\appendix

\clearpage

\begin{minipage}{\textwidth}
{({\bf Affiliations} continued) \\[3mm] \em \footnotesize
$^{21}$South African Large Telescope, P.O. Box 9 Observatory 7935, South Africa\\
$^{22}$Department of Physics and Astrophysics, Faculty of Science, Nagoya
University, Nagoya 464-8602, Japan\\
$^{23}$Armagh Observatory, College Hill, Armagh, BT61 9DG, Northern Ireland,
United Kingdom \\
$^{24}$Korea Astronomy and Space Science Institute, 776 Daedukdae-ro, Yuseong-gu, Daejeon 305-348, Republic of Korea\\
$^{25}$Istituto Internazionale per gli Alti Studi Scientifici (IIASS), Vietri Sul Mare (SA), Italy\\
$^{26}$Max Planck Institute for Astronomy, K\"{o}nigstuhl 17, 619117 Heidelberg, Germany\\
$^{27}$INFN, Gruppo Collegato di Salerno, Sezione di Napoli, Italy\\
$^{28}$Centro de Astro-Ingenier\'ia, Departamento de Astronom\'ia y Astrof\'isica, Pontificia Universidad Cat\'olica de Chile, Casilla 306, Santiago, Chile\\
$^{29}$Deutsches SOFIA Institut, Universit\"{a}t Stuttgart,
Pfaffenwaldring 31, 70569 Stuttgart, Germany\\
$^{30}$SOFIA Science Center, NASA Ames Research Center, Mail Stop
N211-3, Moffett Field CA 94035, United States of America\\
$^{31}$Danmarks Tekniske Universitet, Institut for Rumforskning og -teknologi, Juliane Maries Vej 30, 2100 K¿benhavn, Denmark\\
$^{32}$Bellatrix Astronomical Observatory, Via Madonna de Loco 47,
03023 Ceccano (FR), Italy\\
$^{33}$Department of Physics, Sharif University of Technology,
 P.~O.\ Box 11155--9161, Tehran, Iran \\
$^{34}$Institut d'Astrophysique et de G\'{e}ophysique, All\'{e}e du 6 Ao\^{u}t 17, Sart Tilman, B\^{a}t.\ B5c, 4000 Li\`{e}ge, Belgium\\
$^{35}$European Southern Observatory (ESO)
Alonso de Cordova 3107, Casilla 19001, Santiago 19, Chile \\
$^{36}$Max Planck Institute for Solar System Research, Max-Planck-Str. 2, 37191 Katlenburg-Lindau, Germany\\
$^{37}$Astrophysics Group, Keele University, Staffordshire, ST5 5BG, United Kingdom\\
$^{38}$Astrophysics Research Institute, Liverpool John Moores University, Twelve Quays House, Egerton Wharf, Birkenhead
CH41 1LD, United Kingdom\\
$^{39}$Instituto de Astrof\'{i}sica de Andaluc\'{i}a, Glorieta de la Astronom\'{i}a
s/n, 18008 Granada, Spain\\
$^{40}$Dark Cosmology Centre, Niels Bohr Institutet, K{\o}benhavns Universitet, Juliane Maries Vej 30, 2100 K{\o}benhavn {\O}, Denmark\\
$^{41}$University of Canterbury, Department of Physics and Astronomy, Private Bag 4800, Christchurch 8020, New
Zealand\\
$^{42}$Department of Physics, 225 Nieuwland Science Hall, University of Notre Dame, Notre Dame, IN 46556, United States of America\\
$^{43}$McDonald Observatory, 16120 St Hwy Spur 78 \#2, Fort Davis, TX 79734, United States of America\\
$^{44}$University of Tasmania, School of Mathematics and Physics, Private Bag 37, Hobart, TAS 7001, Australia\\
$^{45}$Institute of Geophysics and Planetary Physics (IGPP), L-413, Lawrence Livermore National Laboratory, PO Box 808, Livermore, CA 94551, United States of America\\
$^{46}$Department of Physics, University of Rijeka, Omladinska 14, 51000 Rijeka, Croatia\\
$^{47}$Technische Universit\"{a}t Wien, Wiedner Hauptstr. 8-10, 1040 Wien, Austria\\
$^{48}$NASA Exoplanet Science Institute, Caltech, MS 100-22, 770 South Wilson Avenue, Pasadena, CA 91125, United States of America\\
$^{49}$Perth Observatory, Walnut Road, Bickley, Perth 6076, WA, Australia\\
$^{50}$South African Astronomical Observatory, P.O. Box 9 Observatory 7935, South Africa\\
$^{51}$Space Telescope Science Institute, 3700 San Martin Drive, Baltimore, MD 21218, United States of America\\
$^{52}$Universidad de Concepci\'{o}n, Departamento de Astronomia,
     Casilla 160--C, Concepci\'{o}n, Chile\\
$^{53}$Department of Physics, Texas A\&M University, 4242 TAMU, College Station, TX 77843-4242, United States of America\\
$^{54}$Institute for Advanced Study, Einstein Drive, Princeton, NJ 08540, United States of America\\
$^{55}$Department of Physics, Ohio State University, 191 W. Woodruff, Columbus, OH 43210, United States of America\\
$^{56}$Mt. John Observatory, P.O. Box 56, Lake Tekapo 8770, New Zealand\\
$^{57}$School of Chemical and Physical Sciences, Victoria University, Wellington, New Zealand\\
$^{58}$Department of Physics, Konan University, Nishiokamoto 8-9-1, Kobe 658-8501, Japan\\
$^{59}$Nagano National College of Technology, Nagano 381-8550, Japan\\
$^{60}$Tokyo Metropolitan College of Aeronautics, Tokyo 116-8523, Japan\\
$^{61}$Department of Earth and Space Science, Osaka University, 1-1 Machikaneyama-cho, Toyonaka, Osaka 560-0043,
Japan\\
$^{62}$Department of Earth Sciences, National Taiwan Normal University, No. 88, Section 4,
Tingzhou Road, Wenshan District, Taipei 11677, Taiwan}\\
\end{minipage}

\end{document}